\documentclass[journal]{new-aiaa}
\usepackage[utf8]{inputenc}
\usepackage{textcomp}

\usepackage{graphicx}
\usepackage{amsmath}
\usepackage[version=4]{mhchem}
\usepackage{siunitx}
\usepackage{longtable,tabularx}
\hypersetup{
	colorlinks=true,
	linkcolor=blue,
	citecolor=blue,
	urlcolor=black
}

\usepackage{bm}
\usepackage{enumerate}
\usepackage{enumitem}
\usepackage{caption}
\usepackage{booktabs}
\usepackage{multirow}
\usepackage{xcolor}
\usepackage{float}
\usepackage{placeins}
\usepackage{threeparttable}
\usepackage{orcidlink}
\usepackage{multicol}
\graphicspath{{myfigures/}} 

\allowdisplaybreaks[4]

\title{Flutter Suppression Enhancement in Coupled Nonlinear Airfoils with Intermittent Mixed Interactions}

\author{Qi Liu\footnote{Postdoctoral Researcher, Department of Systems and Control Engineering; liuqi1780280327@yahoo.com.}\orcidlink{0009-0002-5706-3355}}

\author{Riccardo Muolo\footnote{Postdoctoral Researcher, Department of Systems and Control Engineering; muolo.r.aa@m.titech.ac.jp.}\orcidlink{0000-0001-9093-7478}}

\author{Hiroya Nakao\footnote{Professor, Department of Systems and Control Engineering; nakao@sc.e.titech.ac.jp.}\orcidlink{0000-0003-3394-0392}}
\affil{Institute of Science Tokyo (former Tokyo Tech), 152-8552 Tokyo, Japan}

\author{and \\ Yong Xu\footnote{Professor, School of Mathematics and Statistics; hsux3@nwpu.edu.cn (Corresponding Author).}\orcidlink{0000-0002-8407-4650}}
\affil{Northwestern Polytechnical University, 710072 Xi'an, People's Republic of China}
\affil{MOE Key Laboratory for Complexity Science in Aerospace, Northwestern Polytechnical \\ University, 710072 Xi'an, People's Republic of China}

\begin{document}

\maketitle

\begin{abstract}
Flutter suppression facilitates the improvement of structural reliability to ensure the flight safety of an aircraft. In this study, we propose a novel strategy for enlarging amplitude death (AD) regime to enhance flutter suppression in two coupled identical airfoils with structural nonlinearity. Specifically, we introduce an intermittent mixed coupling strategy, i.e., a linear combination of intermittent instantaneous coupling and intermittent time-delayed coupling between two airfoils. Numerical simulations are performed to reveal the influence mechanisms of different coupling scenarios on the dynamical behaviors of the coupled airfoil systems. The obtained results indicate that the coupled airfoil systems experience the expected AD behaviors within a certain range of the coupling strength and time-delayed parameters. The continuous mixed coupling favors the onset of AD over a larger parameter set of coupling strength than the continuous purely time-delayed coupling. Moreover, the presence of intermittent interactions can lead to a further enlargement of the AD regions, that is, flutter suppression enhancement. Our findings support the structural design and optimization of an aircraft wing for mitigating the unwanted aeroelastic instability behaviors.
\end{abstract}

\section*{Nomenclature}
{\renewcommand\arraystretch{1.0}
\noindent\begin{longtable*}{@{}l @{\quad=\quad} l@{}}
LCOs & limit cycle oscillations \\
AD & amplitude death \\
RMS & root mean square \\
$\alpha$ & pitch angle about the elastic axis for the uncoupled single airfoil \\
$\alpha_i$ & pitch angle about the elastic axis for the coupled $i$-th airfoil \\
$\xi$ & plunge deflection \\
$\xi_i$ & plunge deflection for the coupled $i$-th airfoil \\
$\mu$ & airfoil-air mass ratio \\
$r_{\alpha}$ & radius of gyration about elastic axis \\
$\overline{\omega}$ & ratio of uncoupled natural frequencies of plunge and pitch motion \\
$a_h$ & non-dimensional distance from the mid-chord to the elastic axis \\
$x_{\alpha}$ & non-dimensional distance between the elastic axis and the center of mass \\
$\zeta _{\xi}, \zeta _{\alpha}$ & viscous damping ratios of plunge and pitch motion \\
$\beta_{\xi}, \beta_{\alpha}$ & nonlinear stiffness coefficients of plunge and pitch motion \\
$\phi\left( t \right)$ & Wanger function \\
$\psi_1, \psi_2, \varepsilon_1, \varepsilon_2$ & constants in Wanger function \\
$w_1, w_2, w_3, w_4$ & auxiliary variables for the uncoupled single airfoil \\
$w_{i1}, w_{i2}, w_{i3}, w_{i4}$ & auxiliary variables for the coupled $i$-th airfoil \\
$U^{*}$ & non-dimensional airflow velocity \\
$U_{F}^{*}$ & linear flutter speed \\
$t$ & non-dimensional time \\
$\tau$ & time delay \\
$G\left( \xi \right), M\left( \alpha \right)$ & plunge and pitch stiffness terms\\
$C_L\left( t \right), C_M\left( t \right)$ & aerodynamic lift and pitching moment coefficients \\
$\boldsymbol{x}$ & state vectors for the uncoupled airfoil \\
$\boldsymbol{x}_i$ & state vectors for the coupled $i$-th airfoil \\
$\boldsymbol{F}$ & vector fields for the uncoupled airfoil \\
$\boldsymbol{F}_i$ & vector fields for the coupled $i$-th airfoil \\
$\boldsymbol{p}$ & sets of system parameters for the uncoupled airfoil \\
$\boldsymbol{p}_i$ & sets of system parameters for the coupled $i$-th airfoil \\
$\mathbb{R}$ & set of real numbers \\
$\boldsymbol{x}_{i0}$ & initial values for the coupled $i$-th airfoil \\
$\chi_{T,\theta} \left( t \right)$ & on-off intermittent coupling \\
$n$ & an integer \\
$T$ & on-off period of intermittent coupling \\
$\theta$ & on-off ratio of intermittent coupling \\
$f_{\text{LCOs}}$ & frequency of LCOs \\
$T_{\text{LCOs}}$ & period of LCOs \\
$K$ & coupling strength \\
$\Delta K$ & step size of coupling strength $K$ \\
$\varrho$ & a proportion factor \\
T & transpose operation \\
$x_{\text{RMS}}$ & RMS value of a vibration signal $x\left( t \right)$ \\
$N$ & length of a vibration signal $x\left( t \right)$ \\
\end{longtable*}}

\section{Introduction}

\lettrine{T}{he} problem of aeroelastic instability has become a key limiting factor in the design of modern aircraft structures. Modern aircrafts have unconventional designs, which feature multiple lifting surfaces with different forms of arrangements to improve the aerodynamic performance of the structures. Such configurations are common and have been widely used in practical engineering applications like biplane wings \cite{feng2023control,traub2012experimental,phillips2021design,dhital2022aaeroelastic}, rotating machinery blades \cite{nitti2021spatially}, and box-wing aircraft configuration \cite{gagnon2016aerodynamic,palaia2025metamodeling,russo2020box,salem2025box,gagnon2016euler}. These structures can often be described as multi-plate structures coupled by aerodynamic or structural connections to meet different requirements of aerodynamic performance. To theoretically and numerically explore the aeroelastic behaviors of multiple lifting surface configurations, researchers usually simplify each plate into a two-dimensional airfoil section and structurally couple them to each other via a linear spring \cite{dhital2022baeroelastic,nitti2021spatially}. The aeroelastic properties would be even more complex due to the interaction between structural inertial forces, elastic forces, and aerodynamic forces between the lifting surfaces \cite{lee1999nonlinear}. The aeroelastic behaviors of conceptual airfoil models have attracted widespread interest of scholars in the past decades, especially flutter \cite{lee1999nonlinear,lee2006bifurcation,liu2022complex}. The flutter instability is a critical concern in the design of aircraft structures that can lead to catastrophic oscillations \cite{kong2025dynamics}. Large-amplitude self-sustained limit cycle oscillations (LCOs) are extremely undesired in practical engineering applications, which can pose a deleterious threat to the flight safety and structural integrity of an aircraft. For example, in 1916, an airplane crashed due to the violent tail oscillations caused by flutter \cite{smaili2017intelligent}. As a result, exploring the flutter suppression strategy for wing structures is extremely crucial to ensure the safety and reliability design of an aircraft.

Several studies have been carried out on response prediction and suppression of unwanted vibrations for single airfoil structures. Liu et al. developed several approximated theories like averaging method \cite{liu2023complex,xu2017dynamical} and multi-scale technique \cite{liu2020bistability,ma2022early}, and data-driven approaches like next generation reservoir computing \cite{liu2025reconstructing}, parameter/state estimation \cite{feng2025fusing,liu2021fixed}, and joint noise reduction strategy \cite{yan2025data} to realize response prediction of airfoil structures, which provides the basis for the design and optimization of vibration control. Various active and passive control techniques, including active disturbance rejection \cite{yang2017design}, nonlinear gain-scheduling \cite{lhachemi2017flutter}, linear tuned vibration absorber \cite{verstraelen2017experimental}, and nonlinear energy sinks \cite{lee2007asuppression,lee2007bsuppressing,pidaparthi2019stochastic} have been developed for vibration suppression of wing structures. Liu et al.~\cite{liu2018sliding,liu2018active} developed integral- and fractional-order sliding mode controls to achieve the vibration suppression of conceptual airfoil models with deterministic or random excitations. They also reviewed the latest advances in complex dynamics and vibration suppression of conceptual airfoil models via theoretical, numerical, and data-driven techniques \cite{liu2022complex}. Guo et al.~\cite{guo2023dynamic,guo2024reliability} investigated dynamics and vibration suppression of a hypersonic wing structure in randomly fluctuating flow and improved the reliability by coupling a nonlinear energy sink. Livne \cite{livne2018aircraft} presented an overview of more than 50 years of research in the active flutter control area. However, these works are concerned with the wing structures of conventional aircraft without multiple lifting surface configurations, and achieve the suppression of undesired oscillations through externally control forces or additional vibration absorbers. A remaining open question is whether it is possible to utilize the internal coupling mechanisms between multiple airfoils to achieve the flutter suppression in the wing structures of modern aircraft with multiple airfoils or multiple lifting surface configurations.

From the vast literature on synchronization dynamics \cite{pikovsky1985universal}, we know that the interactions between oscillators, i.e., LCOs, shape the collective behavior and lead to a variety of phenomena, among which collective chaos \cite{nakagawa1993collective,nakao2014complex}, chimera states \cite{kuramoto2002coexistence,zakharova2020chimera,parastesh2021chimeras}, and cluster synchronization \cite{kaneko1990clustering,pecora2014cluster}, just to name a few. One of such synchronization-related behaviors is amplitude death (AD) \cite{saxena2012amplitude,dudkowski2019traveling,zou2023solvable}, which refers to a dynamical phenomenon where the coupled systems are stabilized to a homogeneous steady state (often the system's equilibrium point such as the origin) and the oscillations are suppressed due to the ``death'' of the amplitude (the oscillation amplitude vanishes) as a consequence of coupling. The AD can be used as a potential mechanism for vibration suppression. In recent years, preliminary studies on the AD mechanism have been conducted to achieve the suppression of mechanical structure vibration \cite{dudkowski2019traveling}, thermoacoustic instability \cite{biwa2015amplitude,thomas2018aeffect,thomas2018beffect,srikanth2022self}, and so on. Moreover, some works have begun to explore the suppression of aeroelastic flutter instability using the AD mechanism. For example, Dhital et al.~\cite{dhital2022baeroelastic} investigated the coupling effects of structural and aerodynamic interactions on the aeroelastic behaviors of two airfoils in proximity with instantaneous coupling and also analyzed the effects of coupling stiffness on the oscillation amplitudes. The results show that the pitch and plunge amplitudes of the two airfoils tend to vanish at a certain parameter, which can be regarded as an AD phenomenon. Kirschmeier et al.~\cite{kirschmeier2020amplitude} and Hughes et al.~\cite{hughes2023modulation} investigated dynamics of aeroelastic wing structures undergoing large-amplitude LCOs. They revealed that, under certain conditions, the undesired LCOs were annihilated (i.e., the system returns to equilibrium) due to the aerodynamic coupling between the wing structure and the vortex wake, or the use of a variable-frequency disturbance generator. Raaj et al. explored the AD in coupled identical nonlinear airfoils with time-delayed coupling between pitch angles \cite{raaj2021investigating}, and further discussed the effects of parameter mismatch and time-delayed dissipative coupling on the AD regime of coupled nonlinear airfoils \cite{raj2021effect}. However, these studies on structural coupling focused on either on continuous purely instantaneous coupling or continuous purely time-delayed coupling scenarios. The question of interest is whether a new coupling strategy can be found to further suppress the flutter instability by enhancing the AD regime in coupled airfoils, i.e., expanding the parameter domains where AD occurs.

An interesting framework comes from the intermittent coupling, i.e., a configuration in which the coupling between two LCOs is turned on and off in a periodic or stochastic fashion \cite{ghosh2020comprehending}. The intermittent coupling has been considered in the study of coupled nonlinear oscillators, in the framework of AD \cite{ghosh2022occasional} and synchronization \cite{chen2009synchronization,sun2016theoretical,tian2020symmetry,ghosh2018occasional}. In the presence of intermittent coupling, the AD phenomenon can be induced in a wider range of coupling strength. Ghosh et al.~\cite{ghosh2022occasional} investigated the enhancement of AD regime in typical coupled oscillators, including coupled Stuart--Landau oscillators, Rössler oscillators, and horizontal Rijke tubes, in which intermittent purely time-delayed coupling was considered. They showed that the introduction of intermittent interactions can enhance the AD regime. On the other hand, the coupling signal may be mixed, containing combined effects of instantaneous interaction and time-delayed interaction. Purely instantaneous coupling and purely time-delayed coupling are two limit cases, however, a mixed interaction may be more realistic, reason for which it has been considered in different fields. Zou et al.~\cite{zou2013amplitude} studied the AD phenomenon in coupled nonlinear oscillators with a mixed interaction and revealed that the presence of mixed coupling favors the AD regime over a larger parameter set than the purely instantaneous coupling or purely time-delayed coupling. Bera et al.~\cite{bera2016emergence} explored the AD behaviors in two diffusively coupled Stuart--Landau oscillators with both instantaneous coupling and time-delayed coupling. Nevertheless, the simultaneous effects of the mixed coupling and the intermittent coupling for the suppression of aeroelastic flutter instability of airfoil systems have yet to be determined. Intermittent mixed coupling mechanism, i.e., combining the intermittent instantaneous coupling with intermittent time-delayed coupling, would provide the possibility of enhancing the desired AD regime and further improving the flutter suppression performance. Given its importance for structural design to effectively mitigate the unwanted LCOs in airfoil systems, unveiling the mechanisms behind the onset of AD phenomenon is the main motivation of our investigation, where we aim to enhance AD behaviors in two coupled airfoils via the design of intermittent mixed interactions.

The rest of this paper is organized as follows. Section \ref{sec:sec-2} introduces the definition of a classical on-off intermittent coupling and describes the mathematical model of two coupled airfoils with intermittent mixed interactions. Section \ref{sec:sec-3} reports several numerical experiment results, reveals some interesting phenomena, and analyzes the effects of different coupling scenarios on the AD behaviors. In particular, we give the AD regions of the coupled airfoil systems under the influence of different parameter combinations, as well as numerical statistics on the variation of the length of the AD regions with two key control parameters in the intermittent mixed coupling. Section \ref{sec:sec-4} concludes this paper.

\section{Model} \label{sec:sec-2}

In this Section, we first introduce the intermittent coupling used in this study, and then give the mathematical model of the coupled pitch-plunge airfoils with intermittent mixed interactions, i.e., with the combination of intermittent instantaneous coupling and intermittent time-delayed coupling.

\subsection{Intermittent coupling}

\begin{figure}[!b]
	\centering
	\includegraphics[width=0.48\columnwidth]{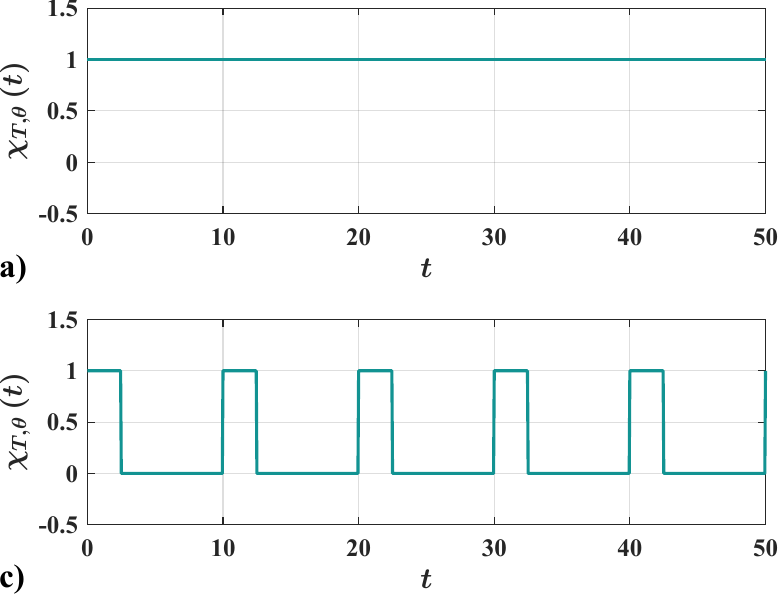} \quad
	\includegraphics[width=0.48\columnwidth]{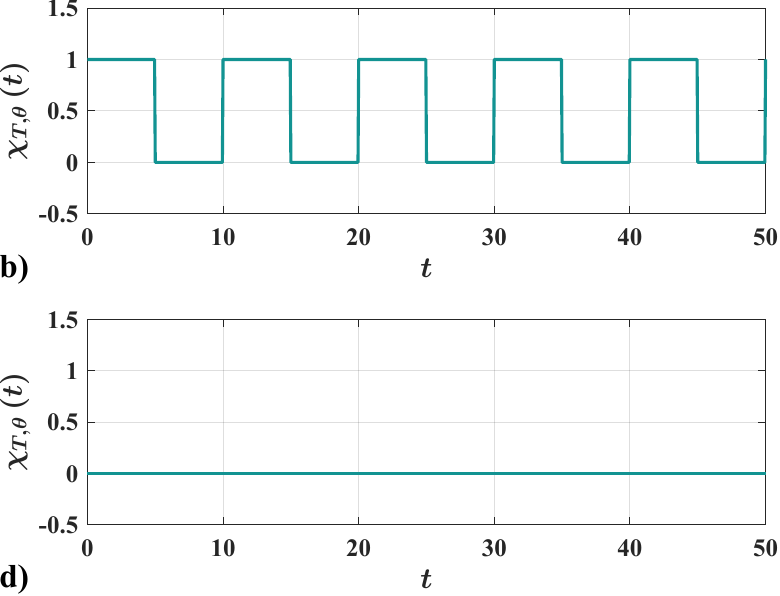}
	\caption{Typical intermittent coupling signals $\chi_{T,\theta} \left( t \right)$ for a fixed on-off period $T=10$ and different on-off ratio $\theta$. (a) $\theta=1$; (b) $\theta=0.50$; (c) $\theta=0.25$; (d) $\theta=0$.}
	\label{fig:intermittent-coupling-signals}
\end{figure}

There are many different types of intermittent coupling, including deterministic and stochastic forms \cite{ghosh2020comprehending}. In the present study, we consider a classical on-off intermittent coupling defined as follows \cite{tian2020symmetry}
\begin{equation} \label{eq:intermittent-coupling}
	\chi_{T,\theta} \left( t \right) =\left\{ \begin{matrix}
		1,&		nT\leq t< \left( n+\theta \right) T,\\
		0,&		\left( n+\theta \right) T\leq t< \left( n+1 \right) T,\\
	\end{matrix} \right.
\end{equation}
in which $n$ is an integer, $T$ is the on-off period, and $\theta \in [0,1]$ is the on-off ratio. The $\chi_{T,\theta} \left( t \right)=1$ represents the state of ``on'', that is, the coupling is active; while the $\chi_{T,\theta} \left( t \right)=0$ indicates the state of ``off'', that is, the coupling is inactive. In order to employ the on-off intermittent coupling, the two control parameters $T$ and $\theta$ should be chosen appropriately, which is a difficult task in practice. According to \cite{ghosh2022occasional}, the average inter-peak separation of the isolated oscillator serves as a reference for appropriate selection of the on-off period $T$. Additionally, the authors of \cite{ghosh2018occasional} pointed out that the on-off period $T$ should be small compared to the system's timescale. However, for the selection of the on-off ratio $\theta$, no such guideline is available. In order to intuitively understand the intermittent coupling mechanism, we show typical signals of the on-off intermittent coupling $\chi_{T,\theta} \left( t \right)$ for a fixed on-off period $T=10$ and different on-off ratio $\theta$, as shown in Fig.~\ref{fig:intermittent-coupling-signals}. We observe that when $\theta=1$, the $\chi_{T,\theta} \left( t \right)=1$ for all $t$, which indicates that the two oscillators are continuously coupled; when $\theta=0$, the $\chi_{T,\theta} \left( t \right)=0$ for all $t$, which represents that the two oscillators are uncoupled; when $\theta\in (0,1)$, the $\chi_{T,\theta} \left( t \right)$ varies periodically between 1 and 0, which means that the two oscillators are intermittently coupled. As a result, Eq.~(\ref{eq:intermittent-coupling}) can be used to model the intermittent interaction between two airfoils.

\subsection{Coupled pitch-plunge airfoils with intermittent mixed interactions}

The multiple lifting surface configurations can be simplified and modeled using conceptual two-dimensional airfoil sections which offer good physical insights into the aeroelastic instability. First, we consider the case of a single airfoil section. In the absence of externally applied forces, the non-dimensional governing equations of a single pitch-plunge airfoil model with nonlinear restoring forces can be expressed as follows \cite{lee1999nonlinear,lee2006bifurcation}

\begin{subequations}
\begin{equation} \label{eq:uncoupled-airfoil-plunge}
	\ddot{\xi}+x_{\alpha}\ddot{\alpha}+2\zeta _{\xi}\frac{\overline{\omega}}{U^*}\dot{\xi}+\left( \frac{\overline{\omega}}{U^*} \right) ^2G\left( \xi \right) =-\frac{1}{\pi \mu}C_L\left( t \right),
\end{equation}
\begin{equation} \label{eq:uncoupled-airfoil-pitch}
	\frac{x_{\alpha}}{r_{\alpha}^{2}}\ddot{\xi}+\ddot{\alpha}+2\zeta _{\alpha}\frac{1}{U^*}\dot{\alpha}+\left( \frac{1}{U^*} \right) ^2M\left( \alpha \right) =\frac{2}{\pi \mu r_{\alpha}^{2}}C_M\left( t \right),
\end{equation}
\end{subequations}
where the dots over the variables represent the derivatives with respect to the non-dimensional time $t$. The $\alpha$ is the pitch angle about the elastic axis which is positive nose up. The $\xi$ is the non-dimensional plunge deflection which is positive in the downward direction. The $x_{\alpha}$ is the non-dimensional distance between the elastic axis and the center of mass. The $r_{\alpha}$ is the radius of gyration about the elastic axis. The $\mu$ is the airfoil–air mass ratio. The $\zeta_{\xi}$ and $\zeta_{\alpha}$ are the viscous damping ratios of plunge and pitch motion, respectively. The $\overline{\omega}$ denotes the frequency ratio of the uncoupled natural frequencies of plunge and pitch motion. The $U^*$ is the non-dimensional airflow velocity. The $G\left( \xi \right)$ and $M\left( \alpha \right)$ represent the linear/nonlinear plunge and pitch stiffness terms, respectively. The $C_L\left( t \right)$ and $C_M\left( t \right)$ are the unsteady aerodynamic lift and pitching moment coefficients, respectively, which are expressed as \cite{lee1999nonlinear,lee2006bifurcation}
\begin{equation*}
	\begin{aligned}
		C_L\left( t \right) &=\pi \left( \ddot{\xi}-a_h\ddot{\alpha}+\dot{\alpha} \right) +2\pi \left[ \alpha \left( 0 \right) +\dot{\xi}\left( 0 \right) +\left( \frac{1}{2}-a_h \right) \dot{\alpha}\left( 0 \right) \right] \phi \left( t \right)\\
		&+2\pi \int_0^{t}{\phi}\left( t -\sigma \right) \left[ \dot{\alpha}\left( \sigma \right) +\ddot{\xi}\left( \sigma \right) +\left( \frac{1}{2}-a_h \right) \ddot{\alpha}\left( \sigma \right) \right] \text{d}\sigma,
	\end{aligned}
\end{equation*}
\begin{equation*}
	\begin{aligned}
		C_M\left( t \right) &=\pi \left( \frac{1}{2}+a_h \right) \left[ \alpha \left( 0 \right) +\dot{\xi}\left( 0 \right) +\left( \frac{1}{2}-a_h \right) \dot{\alpha}\left( 0 \right) \right] \phi \left( t \right)\\
		&+\frac{\pi}{2}\left( \ddot{\xi}-a_h\ddot{\alpha} \right) -\frac{\pi}{16}\ddot{\alpha}-\left( \frac{1}{2}-a_h \right) \frac{\pi}{2}\dot{\alpha}\\
		&+\pi \left( \frac{1}{2}+a_h \right) \int_0^{t}{\phi}\left( t -\sigma \right) \left[ \dot{\alpha}\left( \sigma \right) +\ddot{\xi}\left( \sigma \right) +\left( \frac{1}{2}-a_h \right) \ddot{\alpha}\left( \sigma \right) \right] \text{d}\sigma,
	\end{aligned}
\end{equation*}
in which $a_h$ denotes the non-dimensional distance from the mid-chord to the elastic axis and the Wagner function $\phi \left( t \right) =1-\psi _1e^{-\varepsilon _1t}-\psi _2e^{-\varepsilon _2t}$ with $\psi _1=0.165, \psi _2=0.335, \varepsilon _1=0.0455$ and $\varepsilon _2=0.3$.

In order to deal with the integral terms, we introduce four new variables (augmented states) as follows
\begin{equation*}
	w_{1}=\int_0^{t}{e^{-\varepsilon _1\left( t -\sigma \right)}\alpha \left( \sigma \right) \text{d}\sigma},\quad w_{2}=\int_0^{t}{e^{-\varepsilon _2\left( t -\sigma \right)}\alpha \left( \sigma \right) \text{d}\sigma},
\end{equation*}
\begin{equation*}
	w_{3}=\int_0^{t}{e^{-\varepsilon _1\left( t -\sigma \right)}\xi \left( \sigma \right) \text{d}\sigma},\quad w_{4}=\int_0^{t}{e^{-\varepsilon _2\left( t -\sigma \right)}\xi \left( \sigma \right) \text{d}\sigma},
\end{equation*}
then the airfoil systems (\ref{eq:uncoupled-airfoil-plunge}) and (\ref{eq:uncoupled-airfoil-pitch}) can be rewritten in the form of
\begin{equation*}
	c_0 \ddot{\xi} + c_1 \ddot{\alpha} + c_2 \dot{\xi} + c_3 \dot{\alpha} + c_4 \xi + c_5 \alpha + c_6 w_1 + c_7 w_2 + c_8 w_3 + c_9 w_4 + c_{10} G\left( \xi \right) = f\left( t \right),
\end{equation*}
\begin{equation*}
	d_0 \ddot{\xi} + d_1 \ddot{\alpha} + d_2 \dot{\alpha} + d_3 \alpha + d_4 \dot{\xi} + d_5 \xi + d_6 w_1 + d_7 w_2 + d_8 w_3 + d_9 w_4 + d_{10} M\left( \alpha \right) = g\left( t \right),
\end{equation*}
where the coefficients $c_i$ and $d_i$, $i=0,1,...,10$ depend on the system parameters and are presented in Appendix B, and the $f\left( t \right)$ and $g\left( t \right)$ are expressed as follows
\begin{equation*}
	f\left( t \right) = \frac{2}{\mu} \left[ \left( \frac{1}{2} - a_h \right) \alpha(0) + \xi(0) \right] \left( \psi_1 \varepsilon_1 e^{-\varepsilon_1 t} + \psi_2 \varepsilon_2 e^{-\varepsilon_2 t} \right), \quad g\left( t \right) = -\frac{\left( 1 + 2a_h \right)}{2r_{\alpha}^2} f\left( t \right).
\end{equation*}
When the time $t$ is sufficiently large, i.e., when transients are damped out and steady-state solutions are reached, the function $f\left( t \right) \rightarrow 0$, which implies that also $g\left( t \right) \rightarrow 0$. In the present study, we mainly focus on the steady-state responses of the airfoil systems and set $f\left( t \right)=0$ and $g\left( t \right)=0$ for the sake of simplicity. After introducing a state vector $\boldsymbol{x}=\left( x_{1},x_{2},...,x_{8} \right) ^{\text{T}} =\left( \alpha,\dot{\alpha},\xi,\dot{\xi},w_{1},w_{2},w_{3},w_{4} \right) ^{\text{T}} \in \mathbb{R}^8$, Eqs.~(\ref{eq:uncoupled-airfoil-plunge}) and (\ref{eq:uncoupled-airfoil-pitch}) can be written as
\begin{equation} \label{eq:uncoupled-airfoil-state-equation}
	\dot{\boldsymbol{x}}=\boldsymbol{F}\left( \boldsymbol{x},\boldsymbol{p} \right),
\end{equation}
where $\boldsymbol{p}$ represents the sets of system parameters of the airfoil systems (\ref{eq:uncoupled-airfoil-plunge}) and (\ref{eq:uncoupled-airfoil-pitch}).

Then, we consider the case of two mutually coupled airfoil sections. In the present study, we assume that the two coupled airfoils in proximity are identical and the structural coupling interaction between them is provided through a connection of linear spring \cite{raaj2021investigating}. Besides, we assume that both the airfoils are subjected to the same aerodynamic forces. For two coupled identical airfoils (without parameter mismatch), based on the results \cite{raaj2021investigating}, if both the mixed coupling \cite{zou2013amplitude} and the intermittent coupling \cite{ghosh2022occasional} mechanisms are considered, we obtain the non-dimensional governing equations of two airfoils with intermittent mixed interactions between pitch angles as follows
\begin{subequations}
\begin{equation} \label{eq:coupled-airfoil1-plunge}
	\ddot{\xi}_1+x_{\alpha}\ddot{\alpha}_1+2\zeta _{\xi}\frac{\overline{\omega}}{U^*}\dot{\xi}_1+\left( \frac{\overline{\omega}}{U^*} \right) ^2G\left( \xi _1 \right)=-\frac{1}{\pi \mu}C_L\left( t \right),
\end{equation}
\begin{equation} \label{eq:coupled-airfoil1-pitch}
	\begin{aligned}
		&\frac{x_{\alpha}}{r_{\alpha}^{2}}\ddot{\xi}_1+\ddot{\alpha}_1+2\zeta _{\alpha}\frac{1}{U^*}\dot{\alpha}_1+\left( \frac{1}{U^*} \right) ^2M\left( \alpha _1 \right)+ \left( 1-\varrho \right) \cdot \chi_{T,\theta} \left( t \right) \cdot \frac{K}{U^{*2}} \left( \alpha _1-\alpha _2 \right)\\
		&+ \varrho \cdot \chi_{T,\theta} \left( t \right) \cdot \frac{K}{U^{*2}} \left( \alpha _1-\alpha _2\left( t -\tau \right) \right) =\frac{2}{\pi \mu r_{\alpha}^{2}}C_M\left( t \right),\\
	\end{aligned}
\end{equation}
\end{subequations}
and
\begin{subequations}
\begin{equation} \label{eq:coupled-airfoil2-plunge}
	\ddot{\xi}_2+x_{\alpha}\ddot{\alpha}_2+2\zeta _{\xi}\frac{\overline{\omega}}{U^*}\dot{\xi}_2+\left( \frac{\overline{\omega}}{U^*} \right) ^2G\left( \xi _2 \right)=-\frac{1}{\pi \mu}C_L\left( t \right),
\end{equation}
\begin{equation}
	\label{eq:coupled-airfoil2-pitch}
	\begin{aligned}
		&\frac{x_{\alpha}}{r_{\alpha}^{2}}\ddot{\xi}_2+\ddot{\alpha}_2+2\zeta _{\alpha}\frac{1}{U^*}\dot{\alpha}_2+\left( \frac{1}{U^*} \right) ^2M\left( \alpha _2 \right)+ \left( 1-\varrho \right) \cdot \chi_{T,\theta} \left( t \right) \cdot \frac{K}{U^{*2}} \left( \alpha _2-\alpha _1 \right)\\
		&+ \varrho \cdot \chi_{T,\theta} \left( t \right) \cdot \frac{K}{U^{*2}} \left( \alpha _2-\alpha _1\left( t -\tau \right) \right) =\frac{2}{\pi \mu r_{\alpha}^{2}}C_M\left( t \right),\\
	\end{aligned}
\end{equation}
\end{subequations}
where the subscripts 1 and 2 indicate the first and second airfoils, respectively. The intermittent coupling term $\chi_{T,\theta} \left( t \right)$ is considered as in the form of Eq.~(\ref{eq:intermittent-coupling}). The $K$ is the coupling strength and $\tau$ is the time delay. The parameter $\varrho\in [0,1]$ represents a proportion factor; $\varrho$ is the weight of the time-delayed coupling, and $1-\varrho$ is the weight of the instantaneous coupling. If $\varrho=0$, the coupling signal is purely instantaneous; if $\varrho=1$, the coupling signal is purely time-delayed; and if $\varrho\in (0,1)$, the coupling signal is mixed, i.e., containing both the instantaneous coupling and time-delayed coupling. Eqs.~(\ref{eq:coupled-airfoil1-plunge})-(\ref{eq:coupled-airfoil2-pitch}) can be rewritten in the following state-equation forms of
\begin{subequations}
\begin{equation} \label{eq:coupled-airfoil1-state-equation}
	\dot{\boldsymbol{x}}_1=\boldsymbol{F}_1\left( \boldsymbol{x}_1,\boldsymbol{p} _1 \right) + \left( 1 - \varrho \right) \cdot \chi_{T,\theta} \left( t \right) \cdot \frac{K}{U^{*2}} \left( \boldsymbol{x}_1-\boldsymbol{x}_2 \right) + \varrho \cdot \chi_{T,\theta} \left( t \right) \cdot \frac{K}{U^{*2}} \left( \boldsymbol{x}_1-\boldsymbol{x}_{2\tau} \right),
\end{equation}
\begin{equation} \label{eq:coupled-airfoil2-state-equation}
	\dot{\boldsymbol{x}}_2=\boldsymbol{F}_2\left( \boldsymbol{x}_2,\boldsymbol{p} _2 \right) + \left( 1 - \varrho \right) \cdot \chi_{T,\theta} \left( t \right) \cdot \frac{K}{U^{*2}} \left( \boldsymbol{x}_2-\boldsymbol{x}_1 \right) + \varrho \cdot \chi_{T,\theta} \left( t \right) \cdot \frac{K}{U^{*2}} \left( \boldsymbol{x}_2-\boldsymbol{x}_{1\tau} \right),
\end{equation}
\end{subequations}
where $\boldsymbol{x}_{1\tau}=\boldsymbol{x}_1\left( t-\tau \right)$ and $\boldsymbol{x}_{2\tau}=\boldsymbol{x}_2\left( t-\tau \right)$ denote the time-delayed terms, $\boldsymbol{p}_1$ and $\boldsymbol{p}_2$ are the sets of system parameters of the first and second airfoils, and $\boldsymbol{F}_1\left( \cdot \right)$ and $\boldsymbol{F}_2\left( \cdot \right)$ represent the vector fields. In the following simulations, we assume a constant time delay $\tau$ and initial values $\boldsymbol{x}_1\left( t \right)=\boldsymbol{x}_{10}, \boldsymbol{x}_2\left( t \right)=\boldsymbol{x}_{20}$ for $t \leq 0$. The details are found in Appendix A. It is difficult to perform an analytical study of the stability of the coupled airfoil systems, given that Eqs.~(\ref{eq:coupled-airfoil1-state-equation}) and (\ref{eq:coupled-airfoil2-state-equation}) are coupled to each other, contain time delay, and are high-dimensional. Therefore, we reveal the dynamics of the airfoil systems (\ref{eq:coupled-airfoil1-state-equation})-(\ref{eq:coupled-airfoil2-state-equation}) under different coupling schemes from the perspective of direct numerical simulations.

\section{Results and Discussion} \label{sec:sec-3}

\begin{table}[!b]
	\caption{The values of structural parameters of the aeroelastic airfoil systems (\ref{eq:coupled-airfoil1-state-equation})-(\ref{eq:coupled-airfoil2-state-equation}) \cite{raaj2021investigating}.}
	\centering
	\begin{threeparttable}
		\begin{tabularx}{\textwidth}{lll}
			\toprule
			Parameters & Meaning of symbols & Values \\
			\midrule
			$\mu$ & airfoil-air mass ratio & 555 \\
			$r_{\alpha}$ & radius of gyration about elastic axis & 0.707 \\
			$x_{\alpha}$ & non-dimensional distance between the elastic axis and the center of mass & 0.29 \\
			$a_h$ & non-dimensional distance from the mid-chord to the elastic axis & $-0.5$ \\
			$\overline{\omega}$ & ratio of uncoupled natural frequencies of pitch and plunge motions & 0.99 \\
			$\gamma_1$ & plunge linear stiffness coefficient & 1 \\
			$\gamma_3$ & plunge nonlinear stiffness coefficient & 0 \\
			$\beta_1$ & pitch linear stiffness coefficient & 1 \\
			$\beta_3$ & pitch nonlinear stiffness coefficient & 3 \\
			$\zeta _{\xi}$ & plunge viscous damping ratio & 0.05 \\
			$\zeta _{\alpha}$ & pitch viscous damping ratio & 0.03 \\
			\bottomrule
		\end{tabularx}
	\end{threeparttable}
	\label{tab:values-of-structral-parameters}
\end{table}

In this Section, we will investigate the emergence of AD behaviors in the coupled airfoil systems (\ref{eq:coupled-airfoil1-state-equation})-(\ref{eq:coupled-airfoil2-state-equation}) via direct numerical simulations. We are mainly interested in addressing the three problems: (i) revealing the AD phenomenon in the coupled airfoil systems (\ref{eq:coupled-airfoil1-state-equation})-(\ref{eq:coupled-airfoil2-state-equation}) under the considered coupling effects; (ii) discussing the effects of the parameters $\varrho$, $T$, and $\theta$ on the onset of AD phenomenon; (iii) finding the parameter domains where the expected AD phenomenon can be triggered and comparing the range of these parameters domains for different coupling scenarios.

\subsection{Setup of numerical simulations}

The structural nonlinearities in pitch and plunge degrees of freedom are considered as cubic, that is, $G\left( \xi \right) =\gamma_1 \xi +\gamma_3 \xi ^3$ and $M\left( \alpha \right) =\beta_1 \alpha +\beta_3 \alpha ^3$, where $\gamma_1, \beta_1$ and $\gamma_3, \beta_3$ are the corresponding linear and nonlinear stiffness coefficients of both degrees of freedom, respectively. For the sake of simplicity, we assume that the configurations of the two airfoils are identical (i.e., without parameter mismatch), meaning that $\boldsymbol{p} _{1}=\boldsymbol{p}_{2}$ in Eqs.~(\ref{eq:coupled-airfoil1-state-equation}) and (\ref{eq:coupled-airfoil2-state-equation}). At this point, the two coupled airfoil systems (\ref{eq:coupled-airfoil1-state-equation})-(\ref{eq:coupled-airfoil2-state-equation}) do not show the expected AD regime in the absence of time-delayed coupling, the explanation of which will be given later. The values of structural parameters of the coupled airfoil systems (\ref{eq:coupled-airfoil1-state-equation})-(\ref{eq:coupled-airfoil2-state-equation}) are selected as Table~\ref{tab:values-of-structral-parameters}. The coupled airfoil systems (\ref{eq:coupled-airfoil1-state-equation})-(\ref{eq:coupled-airfoil2-state-equation}) will be solved numerically using the \texttt{ode45} and \texttt{dde23} solvers \cite{shampinea2001solving,shampine2005solving} with default options in the MATLAB R2020a software for the cases without and with time delay, respectively. The initial values of the coupled airfoil systems (\ref{eq:coupled-airfoil1-state-equation})-(\ref{eq:coupled-airfoil2-state-equation}) are chosen as $\boldsymbol{x}_{i0}=\left( 0.5, 0, 0, 0, 0, 0, 0, 0 \right) ^{\text{T}}, i=1, 2$. The values of other parameters will be given below depending on the specific scenario.

\subsection{Bifurcation behaviors of the uncoupled airfoil system (\ref{eq:uncoupled-airfoil-state-equation})}

\begin{figure}[!b]
	\centering
	\includegraphics[width=0.75\columnwidth]{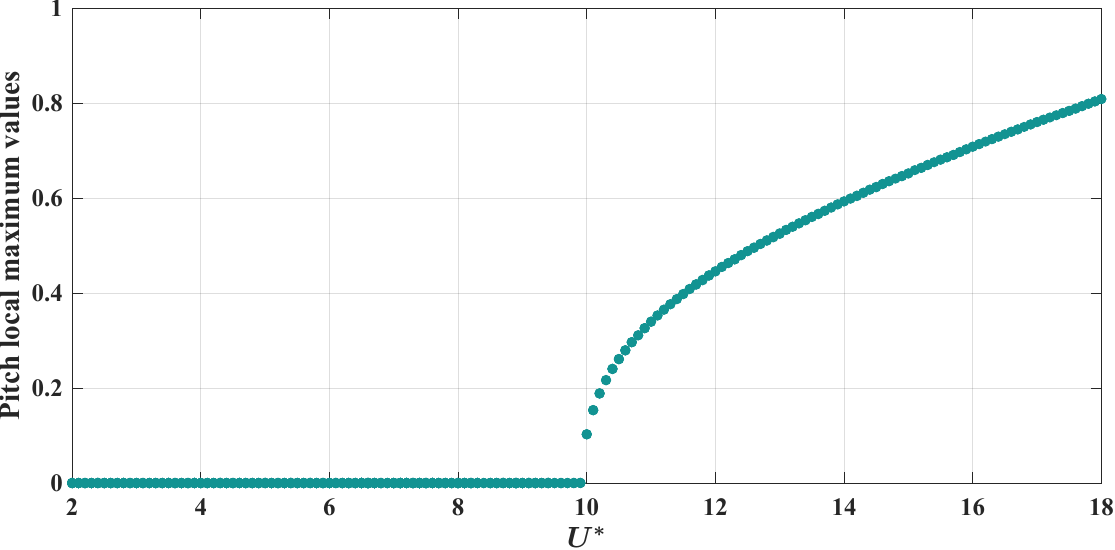}
	\caption{Bifurcation behaviors for the pitch motion of the uncoupled airfoil system (\ref{eq:uncoupled-airfoil-state-equation}).}
	\label{fig:bifurcation-of-uncoupledairfoil}
\end{figure}

\begin{figure}[!t]
	\centering
	\includegraphics[width=0.48\columnwidth]{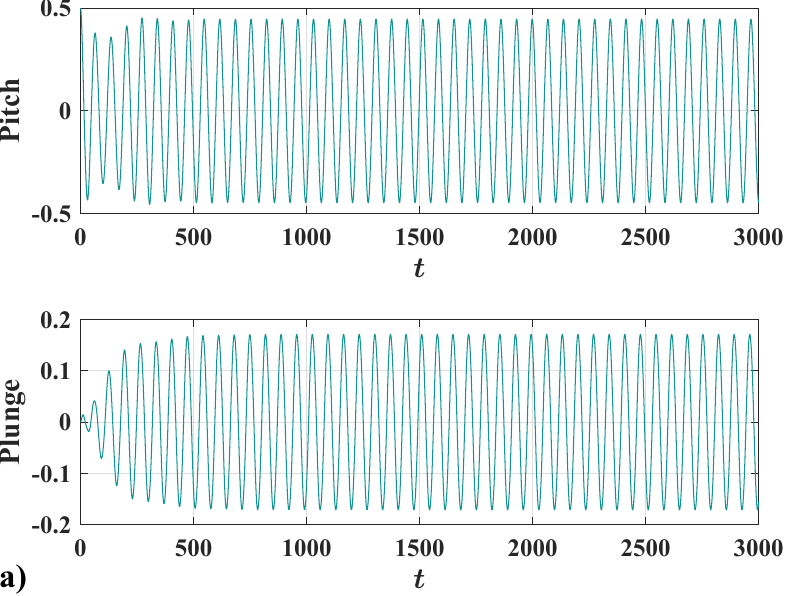} \quad
	\includegraphics[width=0.48\columnwidth]{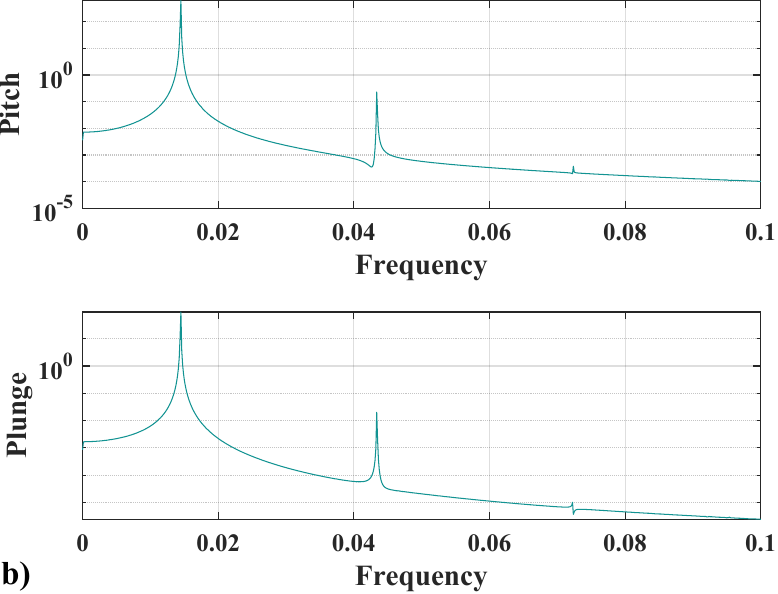}
	\caption{Typical responses of pitch and plunge motions of the uncoupled airfoil system (\ref{eq:uncoupled-airfoil-state-equation}) for the airflow velocity $U^{*}=12$. (a) Time responses; (b) Power spectrum density.}
	\label{fig:TimeseriesandPSD-of-uncoupledairfoil}
\end{figure}

Before considering the effects of the intermittent mixed coupling introduced in the present study, the dynamical behaviors of the uncoupled airfoil system (\ref{eq:uncoupled-airfoil-state-equation}) have to be understood clearly. We first analyze the bifurcation behaviors of the uncoupled airfoil system (\ref{eq:uncoupled-airfoil-state-equation}) with the changing of airflow velocity $U^{*}$, as shown in Fig.~\ref{fig:bifurcation-of-uncoupledairfoil}. The linear flutter velocity $U^{*}_{F}\approx 9.9$. When $U^{*}<U^{*}_{F}$, the uncoupled airfoil system (\ref{eq:uncoupled-airfoil-state-equation}) exhibits typical damped oscillations, that is, converges to a stable zero equilibrium point; however, when $U^{*}>U^{*}_{F}$, the uncoupled airfoil system (\ref{eq:uncoupled-airfoil-state-equation}) exhibits LCOs. Figure \ref{fig:TimeseriesandPSD-of-uncoupledairfoil} displays typical time responses and the corresponding power spectrum density of pitch and plunge motions for the airflow velocity $U^{*}=12$. We find that the uncoupled airfoil system (\ref{eq:uncoupled-airfoil-state-equation}) exhibits typical LCOs whose mean values are zero, and the period of the LCOs is $T_{\text{LCOs}}=1/f_{\text{LCOs}}\approx 68.97$. In practical engineering applications, flutter behavior is usually unpredictable and large-amplitude LCOs can lead to the structural fatigue damage or even threaten the flight safety of an aircraft, which are unexpected and hazardous. Therefore, controlling the undesired LCOs meets the practical requirements and contributes to enhancing the structural reliability of aircraft wings.

\subsection{AD phenomena in the coupled airfoil systems (\ref{eq:coupled-airfoil1-state-equation})-(\ref{eq:coupled-airfoil2-state-equation})}

In this Section, we will discuss in detail the intermittent mixed coupling, i.e., a linear combination of intermittent instantaneous coupling and intermittent time-delayed coupling, on the behavior of the coupled airfoil systems (\ref{eq:coupled-airfoil1-state-equation})-(\ref{eq:coupled-airfoil2-state-equation}). In the present study, we only choose a fixed airflow velocity of $U^{*}=12$ in the flutter region of the uncoupled airfoil system (\ref{eq:uncoupled-airfoil-state-equation}) as an illustrated example. In addition, we only show the results of the first airfoil in the following contents, because the same results can be observed for the second airfoil. We set the simulation time interval $t\in [0,10000]$.

In order to visualize the AD behaviors in the coupled airfoil systems (\ref{eq:coupled-airfoil1-state-equation})-(\ref{eq:coupled-airfoil2-state-equation}), we will use the last 10\% data of a vibration signal to calculate the root mean square (RMS) value of the system responses. For a given discrete vibration signal $\big\{x\left( t_{k} \right)\big\}^{N}_{k=1}$ with $N$ sample points, the RMS value is defined as follows
\begin{equation} \label{eq:RMS-definition}
	x_{\text{RMS}}=\sqrt{\frac{1}{N}\sum_{k=1}^N{x^2\left( t_k \right)}}.
\end{equation}
The RMS value measures the average amplitude (or, physically, the average energy) of a signal over a period of time and is proportional to the amplitude of the oscillations, and it has been widely used for determining AD regime of coupled oscillators \cite{thomas2018aeffect,thomas2018beffect,clusella2021amplitude,srikanth2022self,raaj2021investigating,raj2021effect}. In our system of coupled airfoils (\ref{eq:coupled-airfoil1-state-equation})-(\ref{eq:coupled-airfoil2-state-equation}), when AD occurs, the RMS values $x_{\text{RMS}}$ of both airfoils are close to zero, i.e., $x\left( t \right) \rightarrow 0$. Note that, although the RMS value is zero in our case, in general, it does not necessarily have to be zero, as long as it is homogeneous, i.e., the same for every oscillator. This is different from the case of oscillation death, which is another type of oscillation quenching \cite{koseska2013oscillation}, where the stationary state is inhomogeneous, i.e., not the same for every oscillator. In this study, we are concerned with the suppression of aeroelastic instability in two coupled airfoils, hence, we are only interested in the AD regime. That is, we expect all states of the coupled airfoil systems (\ref{eq:coupled-airfoil1-state-equation})-(\ref{eq:coupled-airfoil2-state-equation}) to be brought to the zero equilibrium state due to our coupling strategy.

In the what follows, we will explore complex dynamical behaviors of the coupled airfoil systems (\ref{eq:coupled-airfoil1-state-equation})-(\ref{eq:coupled-airfoil2-state-equation}) with different coupling scenarios, including: (Case I) continuous purely time-delayed coupling; (Case II) continuous mixed coupling; (Case III) intermittent purely time-delayed coupling; (Case IV) intermittent mixed coupling. We will give the AD regions for different parameter combinations as well as compare the AD regions for different coupling scenarios. Our results illustrate the enhancement of the AD regime due to the proposed coupling strategies.

\subsubsection{Results for the Case I}

\begin{figure}[!t]
	\centering
	\includegraphics[width=0.48\columnwidth]{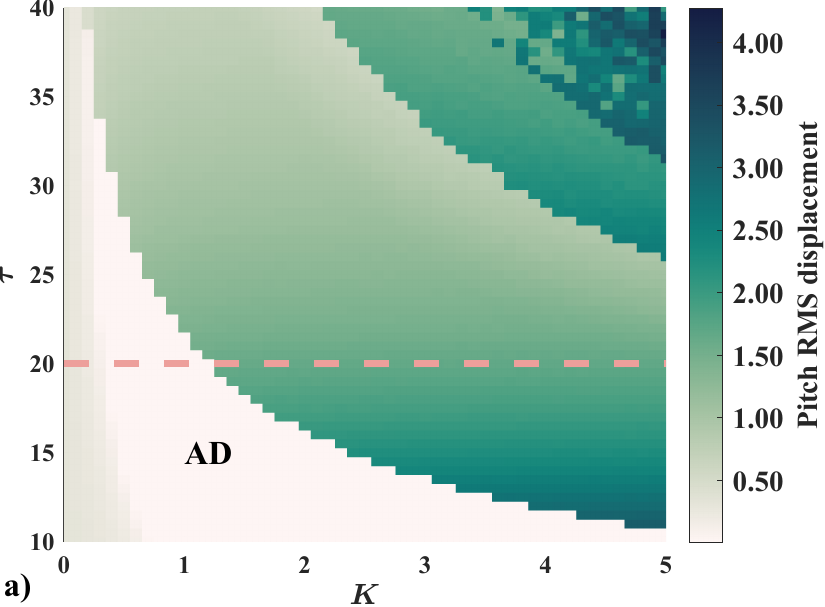} \quad
	\includegraphics[width=0.48\columnwidth]{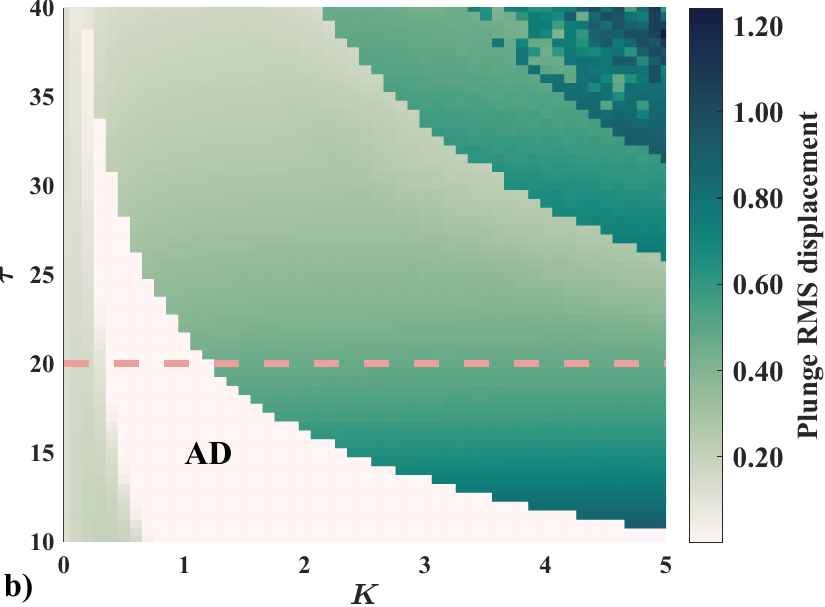}\\
	\caption{The RMS displacements of the coupled airfoil systems (\ref{eq:coupled-airfoil1-state-equation})-(\ref{eq:coupled-airfoil2-state-equation}) for the Case I (i.e., $\varrho=1$ and $\chi_{T,\theta} \left( t \right)\equiv1$). (a) Plunge motion; (b) Plunge motion.}
	\label{fig:RMS-displacements-and-bifurcation-CaseI-Ktau}
\end{figure}

\begin{figure}[!t]
	\centering
	\includegraphics[width=0.75\columnwidth]{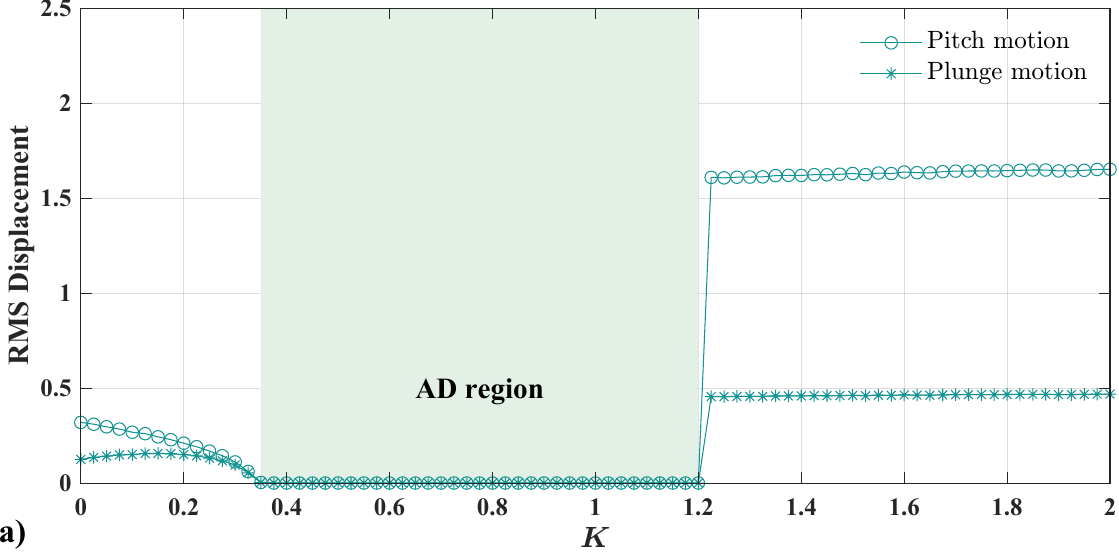} \\
	\includegraphics[width=0.75\columnwidth]{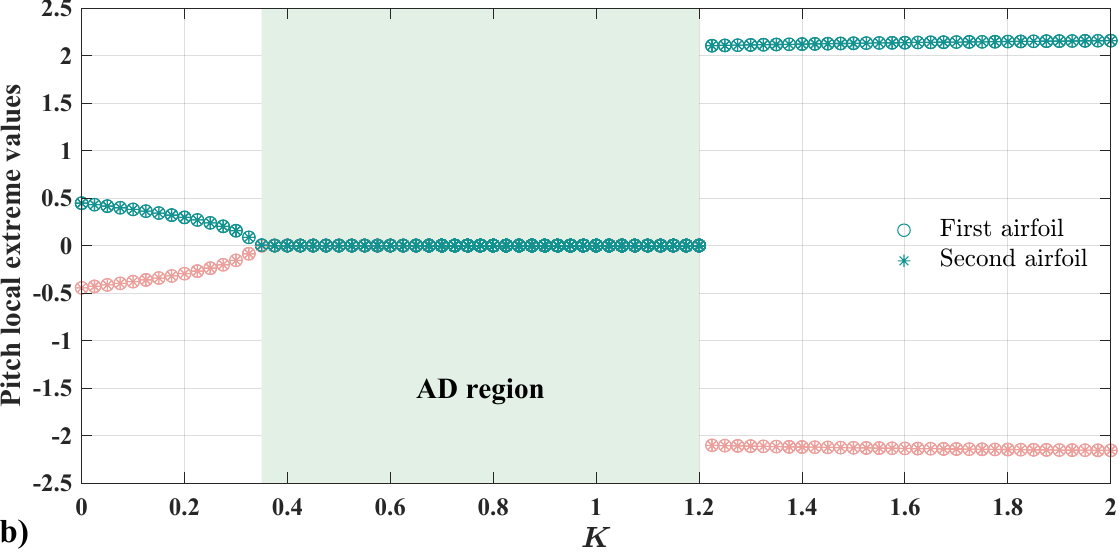}
	\caption{The results of the coupled airfoil systems (\ref{eq:coupled-airfoil1-state-equation})-(\ref{eq:coupled-airfoil2-state-equation}) for the Case I under $\tau=20$. (a) RMS displacements of the first airfoil versus $K$; (b) Bifurcation behaviors of the first and second airfoils versus $K$ (green marker: local maximum values, light red marker: local minimum values).}
	\label{fig:RMS-displacements-and-bifurcation-CaseI-K}
\end{figure}

\begin{figure}[!t]
	\centering
	\includegraphics[width=0.48\columnwidth]{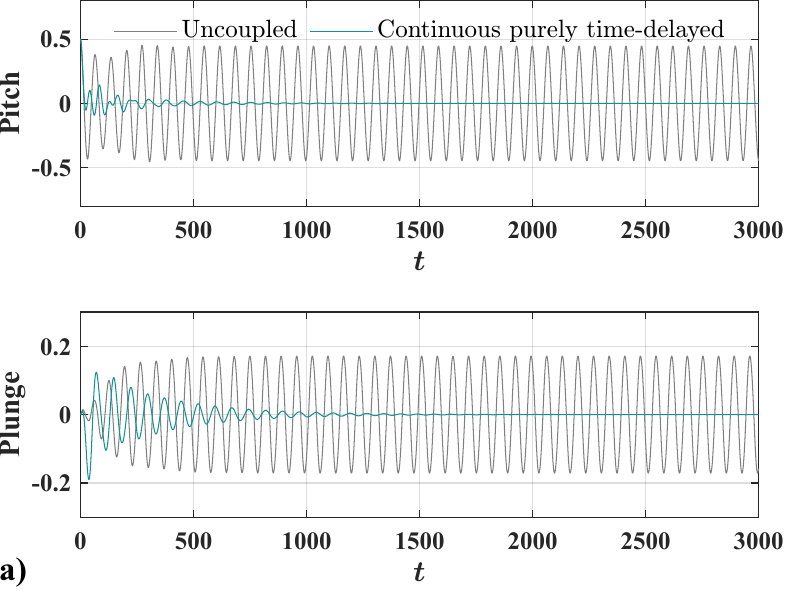} \quad
	\includegraphics[width=0.48\columnwidth]{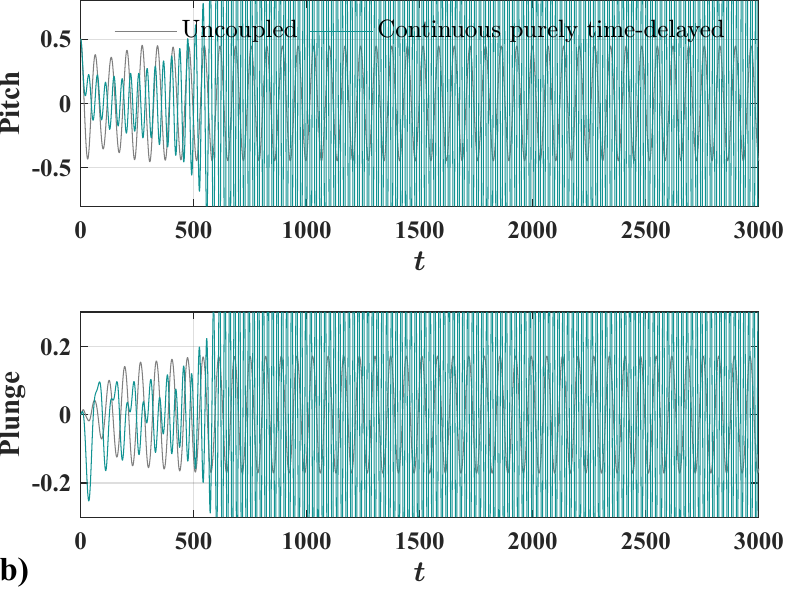}
	\caption{Typical time responses of the coupled airfoil systems (\ref{eq:coupled-airfoil1-state-equation})-(\ref{eq:coupled-airfoil2-state-equation}) for the Case I under a fixed time delay $\tau=20$ and different coupling strengths $K$. (a) $K=0.8$; (b) $K=1.3$.}
	\label{fig:time-series-CaseI-K}
\end{figure}

In order to highlight the advantages of the proposed new coupling strategy, we first study the degenerate case where the coupled airfoil systems (\ref{eq:coupled-airfoil1-state-equation})-(\ref{eq:coupled-airfoil2-state-equation}) contains only the continuous purely time-delayed coupling without intermittent interactions, i.e., $\varrho=1$ and $\chi_{T,\theta} \left( t \right)\equiv 1$. According to the definition of Eq.~(\ref{eq:RMS-definition}), we calculate the RMS displacements of pitch and plunge motions of the coupled airfoil systems (\ref{eq:coupled-airfoil1-state-equation})-(\ref{eq:coupled-airfoil2-state-equation}) in the parameter space $(K, \tau)$. Figure \ref{fig:RMS-displacements-and-bifurcation-CaseI-Ktau} only shows the RMS displacements of the first airfoil. For the sake of brevity, the second airfoil exhibits similar responses, which will not be presented in this paper. The parameter space we consider is $(K, \tau)=[0,5] \times [10,40]$. The white regions indicate the parameter domains where the coupled airfoil systems (\ref{eq:coupled-airfoil1-state-equation})-(\ref{eq:coupled-airfoil2-state-equation}) will experience AD. We find that for some suitable combinations of parameters, the coupled airfoil systems (\ref{eq:coupled-airfoil1-state-equation})-(\ref{eq:coupled-airfoil2-state-equation}) exhibit the AD behaviors. That is, the system oscillations will converge to a stable zero equilibrium point and the unwanted LCOs are effectively controlled. In addition, as the time delay $\tau$ gradually increases, the range of the coupling strength $K$ leading to AD will gradually decrease. In other words, the small time delay $\tau$ is more effective in achieving the purpose of flutter suppression.

To visualize the effects of the coupling strength $K$ on the AD behaviors more intuitively, we present the effects of the coupling strength $K$ on the dynamical behaviors of the coupled airfoil systems (\ref{eq:coupled-airfoil1-state-equation})-(\ref{eq:coupled-airfoil2-state-equation}) for a fixed time delay $\tau=20$ (corresponding to the light red dashed line in Fig.~\ref{fig:RMS-displacements-and-bifurcation-CaseI-Ktau}), as shown in Fig.~\ref{fig:RMS-displacements-and-bifurcation-CaseI-K}. The results of Figs.~\ref{fig:RMS-displacements-and-bifurcation-CaseI-K}a and \ref{fig:RMS-displacements-and-bifurcation-CaseI-K}b demonstrate the variation of the RMS displacements of the first airfoil and the bifurcation behaviors of the first and second airfoils with the coupling strength $K$ in the stationary regime, respectively. The step size of $K$ is chosen as $\Delta K=0.025$. The green area represents the AD region in the coupled airfoil systems (\ref{eq:coupled-airfoil1-state-equation})-(\ref{eq:coupled-airfoil2-state-equation}). In Figs.~\ref{fig:RMS-displacements-and-bifurcation-CaseI-K}b, the circle and star symbols represent the results of the first and second airfoils, respectively. The green and light red markers indicate the local maximum and minimum values of the pitch responses of the coupled airfoil systems (\ref{eq:coupled-airfoil1-state-equation})-(\ref{eq:coupled-airfoil2-state-equation}), respectively.

We can observe from Figs.~\ref{fig:RMS-displacements-and-bifurcation-CaseI-K}a that, as the coupling strength $K$ increases, the RMS displacement of pitch motion gradually decreases, while the RMS displacement of plunge motion first increases and then decreases. Furthermore, the RMS displacements of both pitch and plunge motions are close to zero, that is, the AD occurs, as $K$ increases to a critical value. The AD phenomenon disappears and the RMS displacements of pitch and plunge motions gradually increase as $K$ increases further beyond a certain critical value. Incidentally, plotting the variation diagram of the RMS displacements using any of the remaining variables yields the same conclusion. From Fig.~\ref{fig:RMS-displacements-and-bifurcation-CaseI-K}b, we can observe that since the two coupled airfoils are identical without parameter mismatch, the dynamical behaviors of the first and second airfoils are the same under the coupling effect. For a fixed coupling strength $K$ in the considered parameter range, there exists only one local maximum value and one local minimum value for each airfoil system. Outside the green region, the local maximum and minimum values of pitch responses are distinct and symmetric, indicating that the coupled airfoil systems (\ref{eq:coupled-airfoil1-state-equation})-(\ref{eq:coupled-airfoil2-state-equation}) are in oscillatory states. As the coupling strength $K$ gradually increases from zero, the local maximum and minimum values of pitch responses gradually approach each other. When the coupling strength $K$ exceeds a critical value to reach the green area, the local maximum and minimum values of pitch responses will be equal and zero, indicating that the coupled airfoil systems (\ref{eq:coupled-airfoil1-state-equation})-(\ref{eq:coupled-airfoil2-state-equation}) are stabilized to the same (homogeneous) steady state (zero equilibrium point). In other words, the expected AD regime occurs. The results confirm the correctness of the identified AD region marked in Fig.~\ref{fig:RMS-displacements-and-bifurcation-CaseI-K}a, and verify the applicability and effectiveness of the RMS displacement for determining the AD regimes of the coupled airfoil systems (\ref{eq:coupled-airfoil1-state-equation})-(\ref{eq:coupled-airfoil2-state-equation}) considered in this study.

Additionally, we plot the time responses of the coupled airfoil systems (\ref{eq:coupled-airfoil1-state-equation})-(\ref{eq:coupled-airfoil2-state-equation}) for two typical coupling strengths $K=0.8$ and $K=1.3$ under a fixed time delay $\tau=20$ in Fig.~\ref{fig:time-series-CaseI-K}. Here we only present the results of the first airfoil. It can be observed that when $K=0.8$, the coupled airfoil systems (\ref{eq:coupled-airfoil1-state-equation})-(\ref{eq:coupled-airfoil2-state-equation}) experience AD, i.e., the LCOs converge to the stable zero equilibrium point. Although not shown here, we also numerically confirmed that other states of the coupled airfoil systems (\ref{eq:coupled-airfoil1-state-equation})-(\ref{eq:coupled-airfoil2-state-equation}) exhibit similar AD phenomena. All the system states will be stabilized to the desired zero equilibrium point, i.e., the aeroelastic flutter instability is effectively suppressed. However, when $K=1.3$, the coupled airfoil systems (\ref{eq:coupled-airfoil1-state-equation})-(\ref{eq:coupled-airfoil2-state-equation}) will experience larger-amplitude LCOs than the uncoupled case. These results suggest that we should appropriately choose the coupling strength $K$ to achieve the purpose of flutter suppression.

\subsubsection{Results for the Case II}

\begin{figure}[!t]
	\centering
	\includegraphics[width=0.48\columnwidth]{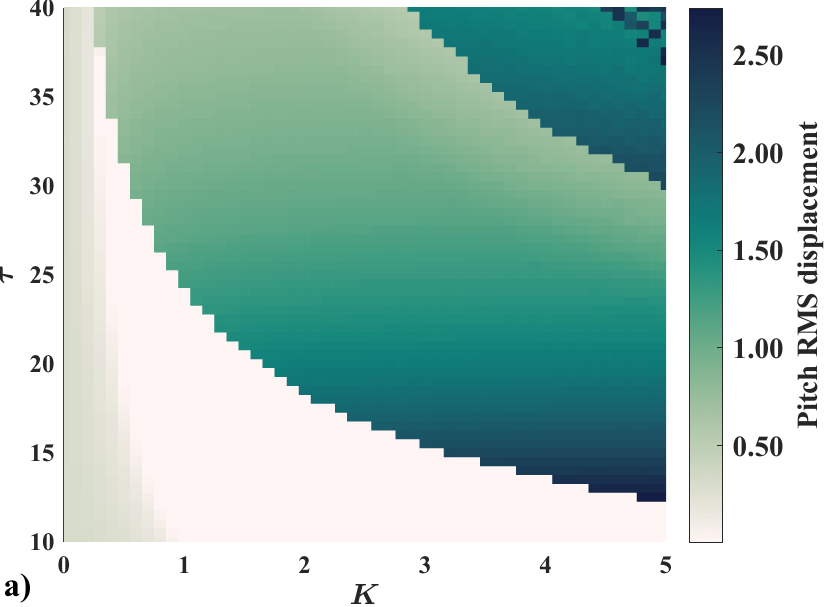} \quad
	\includegraphics[width=0.48\columnwidth]{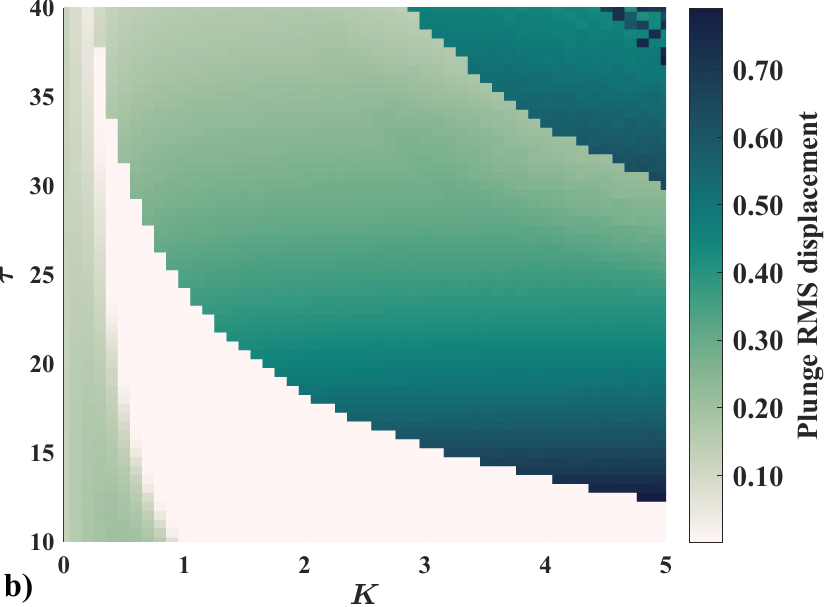} \\
	\includegraphics[width=0.48\columnwidth]{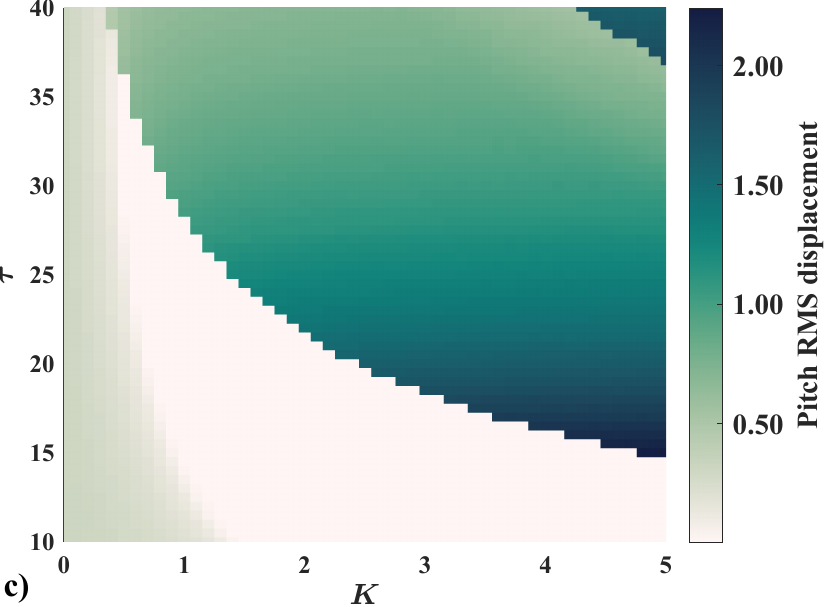} \quad
	\includegraphics[width=0.48\columnwidth]{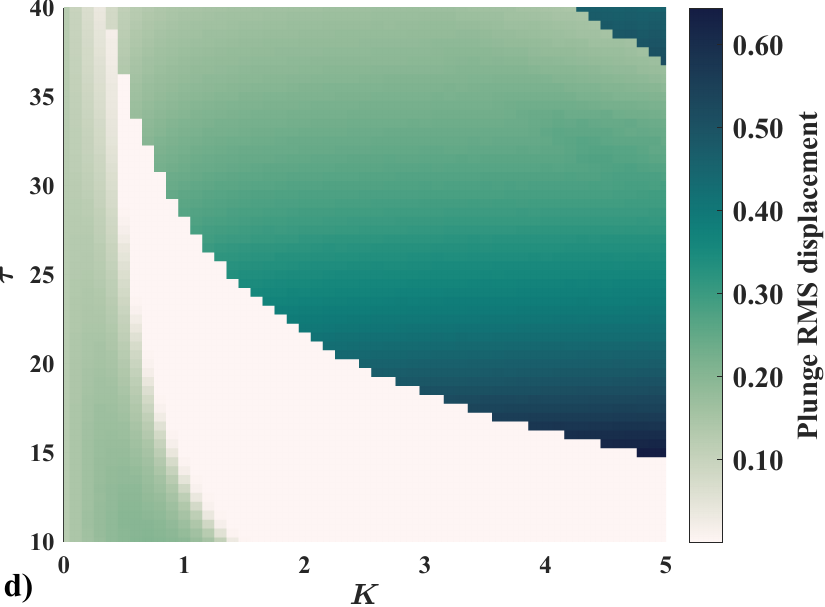} \\
	\includegraphics[width=0.48\columnwidth]{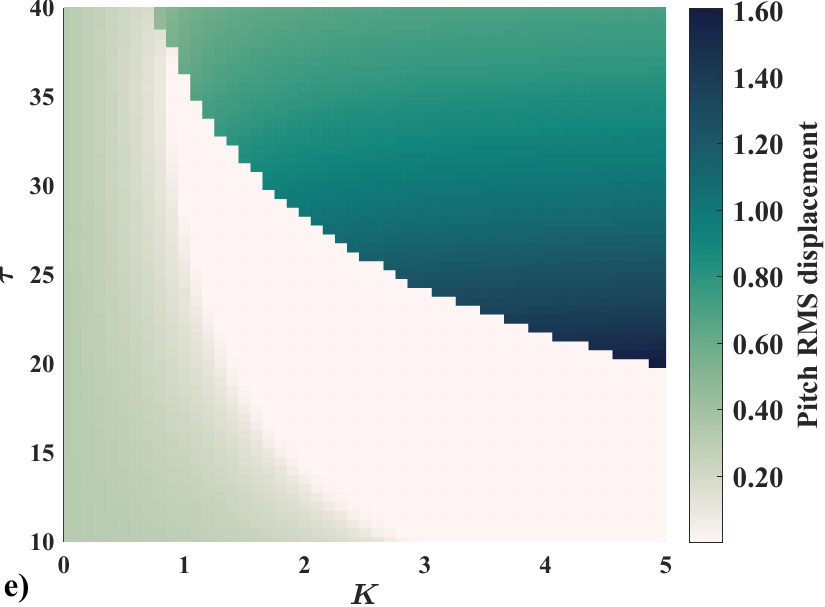} \quad
	\includegraphics[width=0.48\columnwidth]{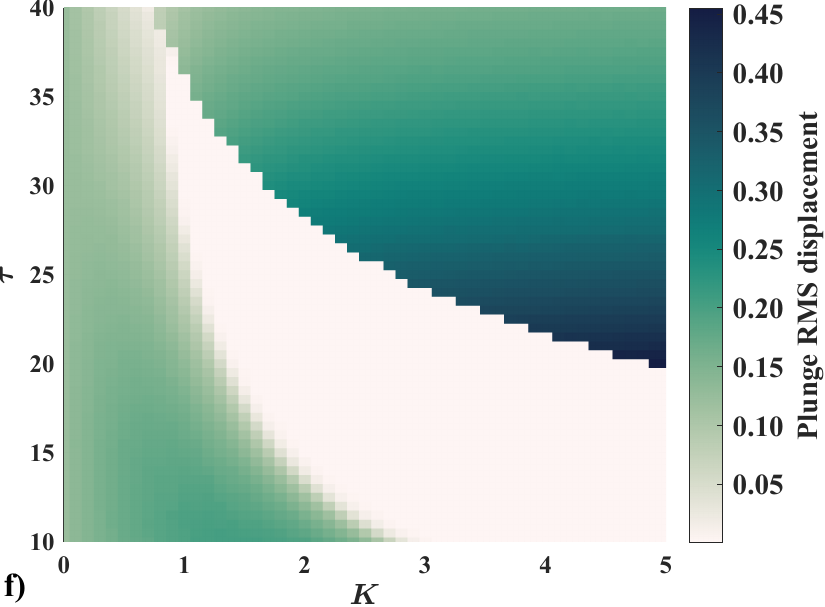}
	
	\caption{The RMS displacements of the coupled airfoil systems (\ref{eq:coupled-airfoil1-state-equation})-(\ref{eq:coupled-airfoil2-state-equation}) for the Case II under different coupling strengths $K$ and time delays $\tau$. (a, b) $\varrho=0.75$; (c, d) $\varrho=0.5$; (e, f) $\varrho=0.25$.}
	\label{fig:RMS-displacements-CaseII-Ktau-varrho}
\end{figure}

\begin{figure}[!t]
	\centering
	\includegraphics[width=0.48\columnwidth]{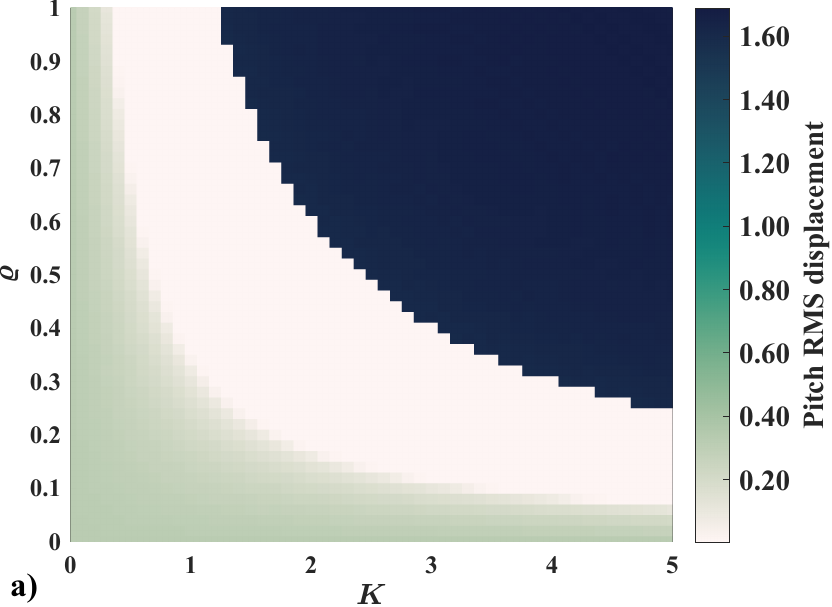} \quad
	\includegraphics[width=0.48\columnwidth]{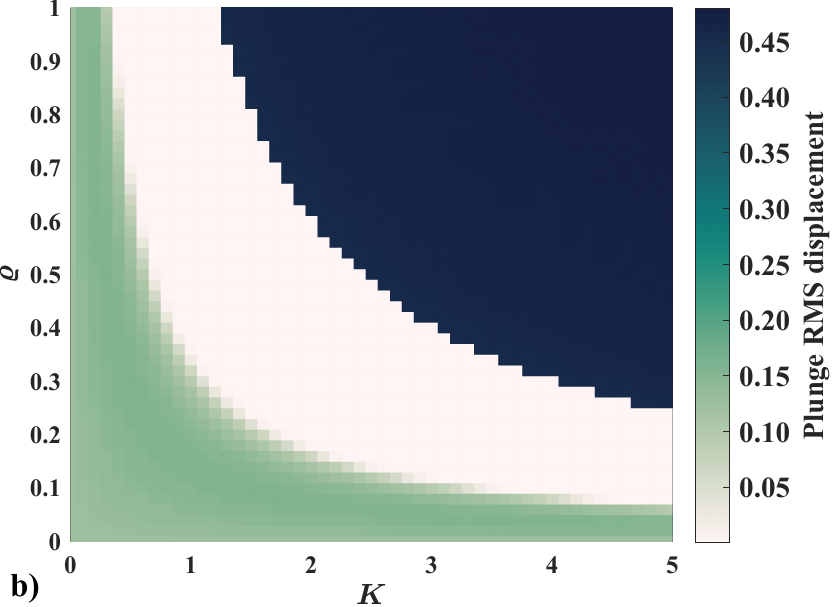} \\
	\includegraphics[width=0.48\columnwidth]{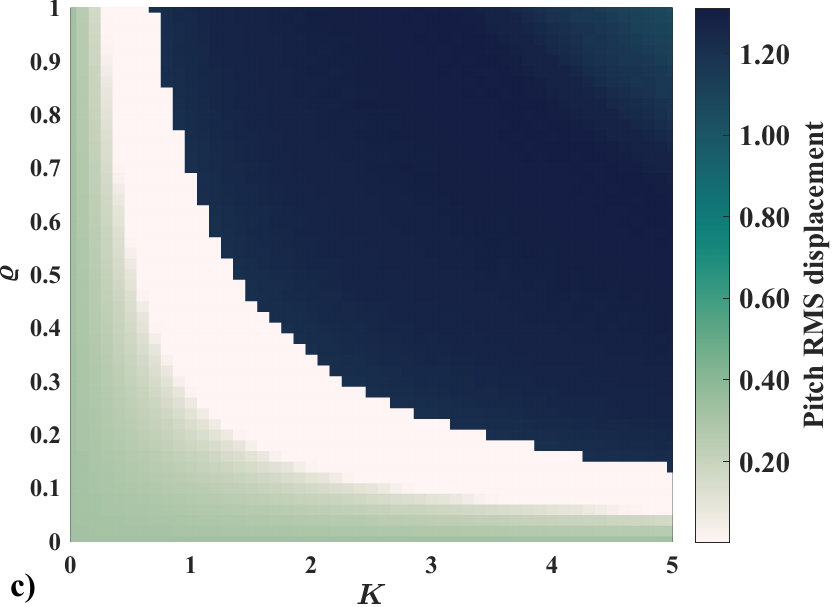} \quad
	\includegraphics[width=0.48\columnwidth]{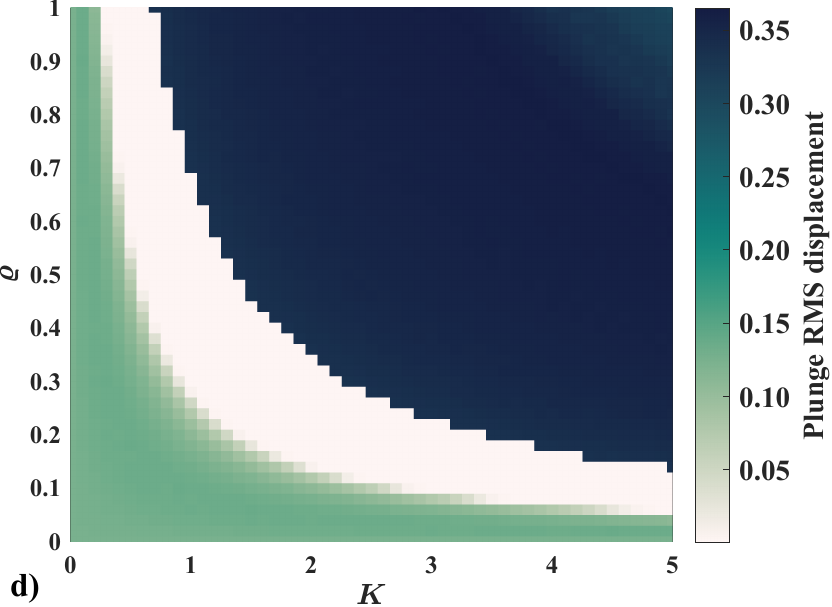} \\
	\includegraphics[width=0.48\columnwidth]{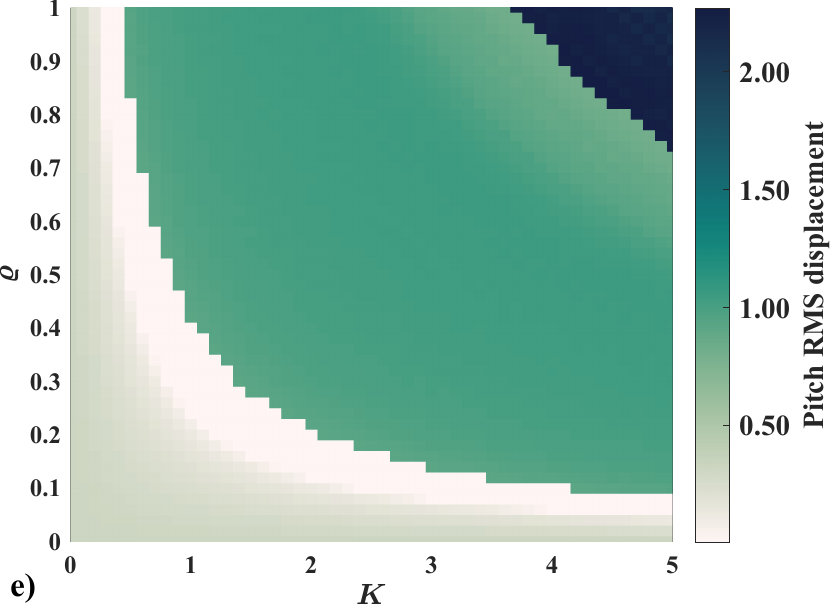} \quad
	\includegraphics[width=0.48\columnwidth]{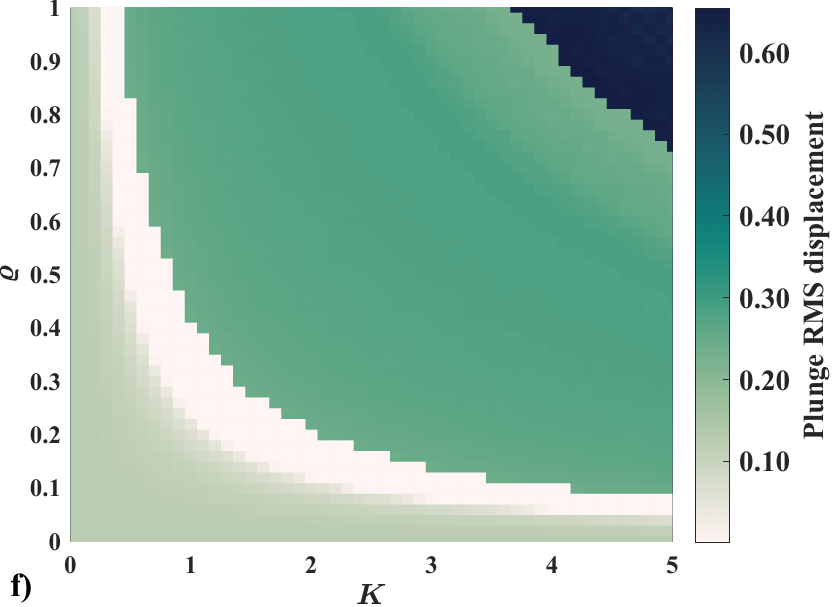}
	\caption{The RMS displacements of the coupled airfoil systems (\ref{eq:coupled-airfoil1-state-equation})-(\ref{eq:coupled-airfoil2-state-equation}) for the Case II under different coupling strengths $K$ and parameters $\varrho$. (a, b) $\tau=20$; (c, d) $\tau=25$; (e, f) $\tau=30$.}
	\label{fig:RMS-displacements-CaseII-Kvarrho-tau}
\end{figure}

\begin{figure}[!t]
	\centering
	\includegraphics[width=0.48\columnwidth]{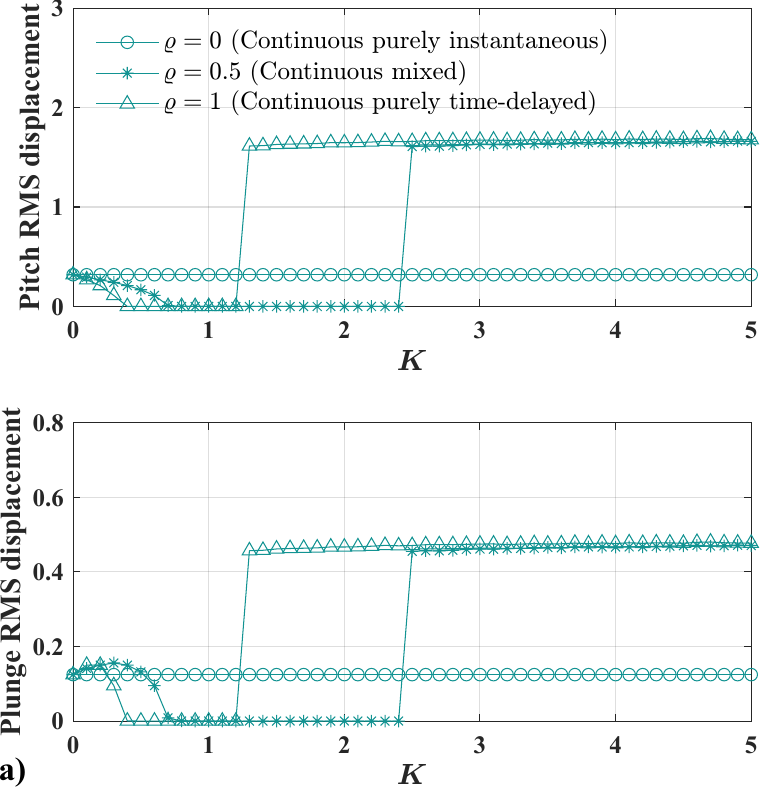} \quad
	\includegraphics[width=0.48\columnwidth]{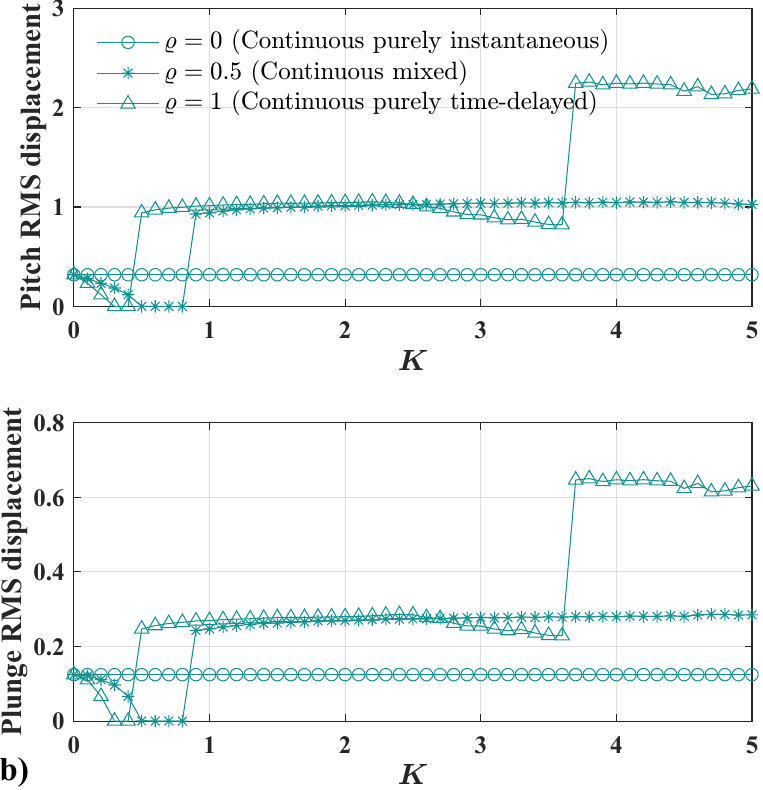}
	\caption{The effects of the parameter $\varrho$ on the RMS displacements of the coupled airfoil systems (\ref{eq:coupled-airfoil1-state-equation})-(\ref{eq:coupled-airfoil2-state-equation}) for the Case II under different time delay $\tau$. (a) $\tau=20$; (b) $\tau=30$.}
	\label{fig:RMS-displacements-CaseII-K-tau}
\end{figure}

Next, we consider the dynamical behaviors of the coupled airfoil systems (\ref{eq:coupled-airfoil1-state-equation})-(\ref{eq:coupled-airfoil2-state-equation}) with continuous mixed coupling, i.e., the combination of continuous instantaneous coupling and continuous time-delayed coupling. In this case, the parameter $\varrho\in (0,1)$ and $\chi_{T,\theta} \left( t \right)\equiv 1$, and discuss the effects of different values of $\varrho$. We plot the RMS displacements of the coupled airfoil systems (\ref{eq:coupled-airfoil1-state-equation})-(\ref{eq:coupled-airfoil2-state-equation}) under different combinations of coupling strengths $K$ and time delay $\tau$ for different values of $\varrho$, including $\varrho=0.75$, $\varrho=0.50$, and $\varrho=0.25$, respectively, as shown in Fig.~\ref{fig:RMS-displacements-CaseII-Ktau-varrho}. The results show that the coupled airfoil systems (\ref{eq:coupled-airfoil1-state-equation})-(\ref{eq:coupled-airfoil2-state-equation}) exhibit AD under the effects of coupling, and the range of coupling strengths $K$ in which such AD phenomenon occurs gradually increases as $\varrho$ decreases.

To clearly visualize the effects of $\varrho$, we present the results in the parameter space ($K, \varrho$) for different time delays $\tau$, as shown in Fig.~\ref{fig:RMS-displacements-CaseII-Kvarrho-tau}. The parameter space we consider is $(K, \varrho)=[0,5] \times [0,1]$. The results show that, as $\varrho$ decreases, the parameter range of coupling strength $K$ in which AD occurs increases, but the minimum critical value of the coupling strength $K$ that leads to AD gradually increases. At the same time, we find that when the continuous mixed coupling degrades to the continuous purely instantaneous coupling (i.e., $\varrho=0$), the coupled airfoil systems (\ref{eq:coupled-airfoil1-state-equation})-(\ref{eq:coupled-airfoil2-state-equation}) do not present the AD regime due to the absence of parameter mismatch and time delay. In other words, AD does not occur in a system of two identical oscillators without time-delayed coupling. For standard symmetric, instantaneous, and linear coupling, without parameter mismatch and time delay, coupled multiple oscillators tend to be synchronized, either in phase or in anti-phase, but without spontaneously entering a steady state (i.e., AD never occurs). Triggering the AD regime usually requires the introduction of mechanisms such as parameter mismatch, time delay, or asymmetric coupling, which has been proven for various systems of coupled nonlinear oscillators \cite{aronson1990amplitude,reddy1998time,teki2017amplitude,saxena2025revisiting}.

Furthermore, we show the effects of the parameter $\varrho$ on the RMS displacements of the coupled airfoil systems (\ref{eq:coupled-airfoil1-state-equation})-(\ref{eq:coupled-airfoil2-state-equation}) for Case II under fixed time delays $\tau=20$ and $\tau=30$, respectively, as indicated in Fig.~\ref{fig:RMS-displacements-CaseII-K-tau}. We show only three typical cases, including the continuous purely instantaneous coupling ($\varrho=0$), continuous mixed coupling ($\varrho=0.5$), and continuous purely time-delayed coupling ($\varrho=1$). The results indicate that for two identical airfoils, the coupled airfoil systems (\ref{eq:coupled-airfoil1-state-equation})-(\ref{eq:coupled-airfoil2-state-equation}) do not experience the AD phenomenon for the case of continuous purely instantaneous coupling and the RMS displacements remain essentially constant as the coupling strength $K$ increases. When considering the continuous mixed coupling, the coupled airfoil systems (\ref{eq:coupled-airfoil1-state-equation})-(\ref{eq:coupled-airfoil2-state-equation}) appear AD for a certain range of the coupling strength $K$. Let us point out that, compared to the continuous purely time-delayed coupling, the continuous mixed coupling can enhance the AD regime, i.e., expanding the parameter range of coupling strengths $K$ in which AD occurs, and thus improve the flutter suppression performance. Moreover, the performance in enhancing the desired AD regimes is weaker for large time delays $\tau$ than that for small time delays $\tau$. However, large time delays $\tau$ can suppress the amplitudes of oscillations when the coupling strength $K$ is beyond the critical value where the AD phenomenon disappears.

\subsubsection{Results for the Case III}

\begin{figure}[!t]
	\centering
	\includegraphics[width=0.48\columnwidth]{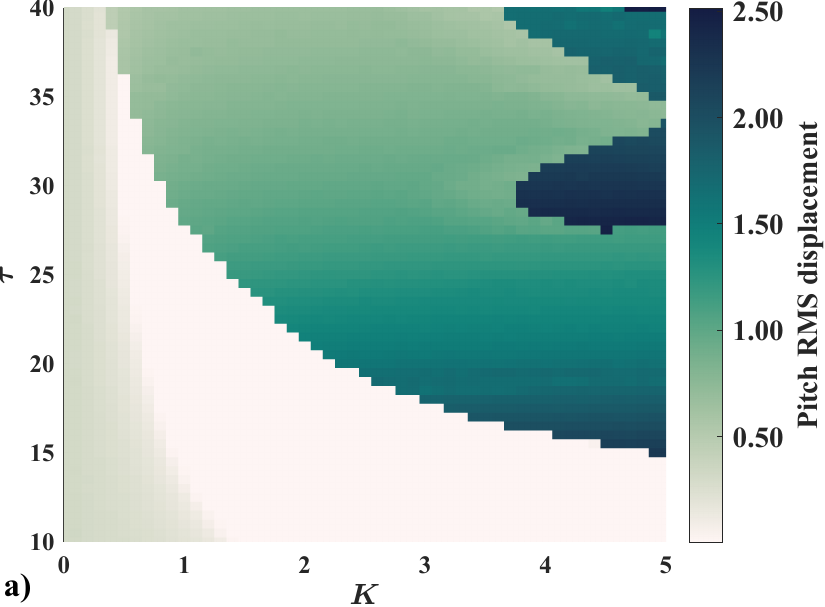} \quad
	\includegraphics[width=0.48\columnwidth]{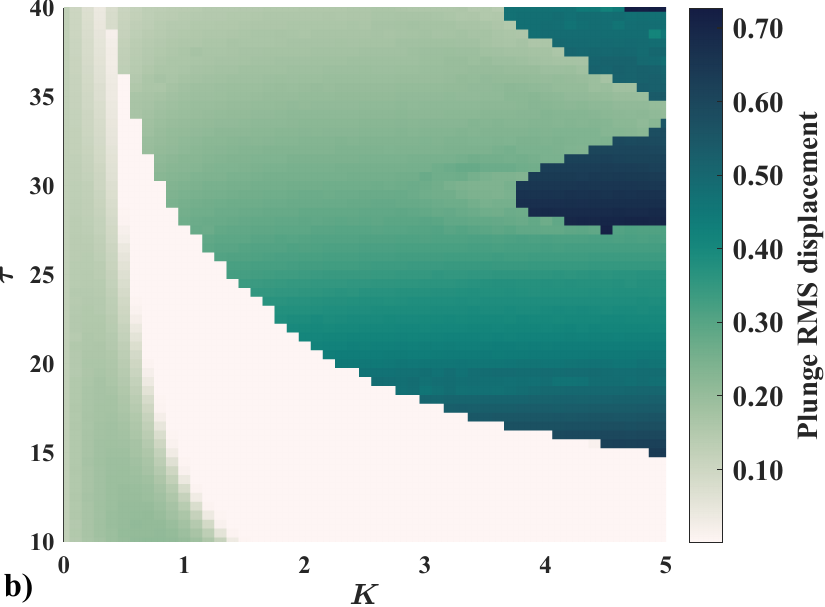} \\
	\includegraphics[width=0.48\columnwidth]{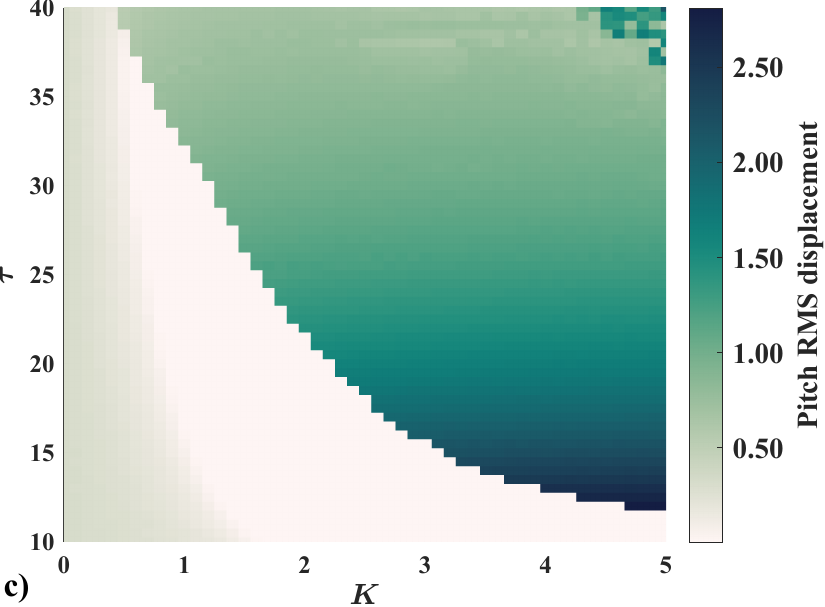} \quad
	\includegraphics[width=0.48\columnwidth]{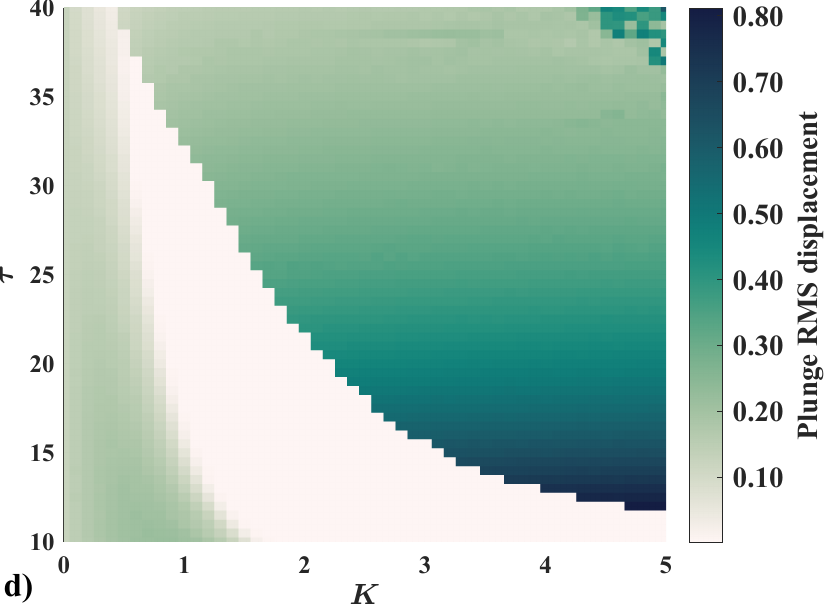} \\
	\includegraphics[width=0.48\columnwidth]{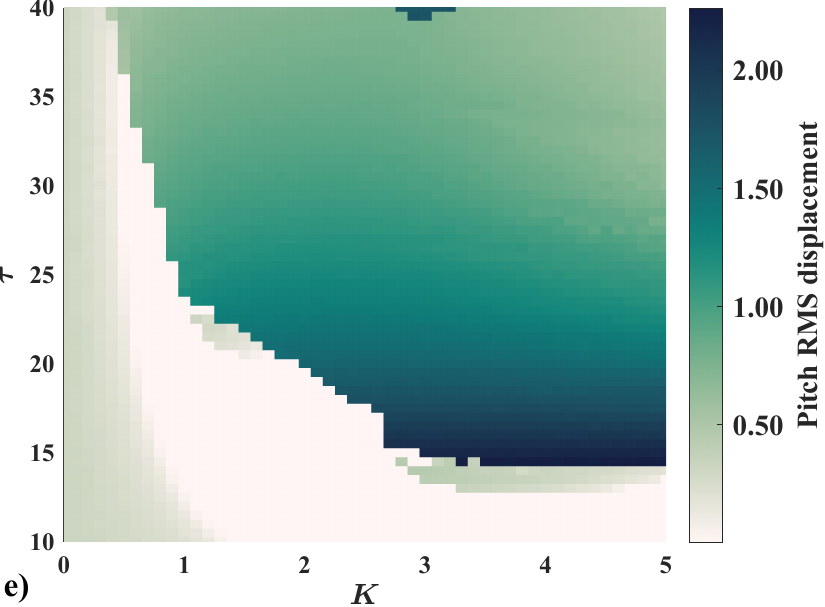} \quad
	\includegraphics[width=0.48\columnwidth]{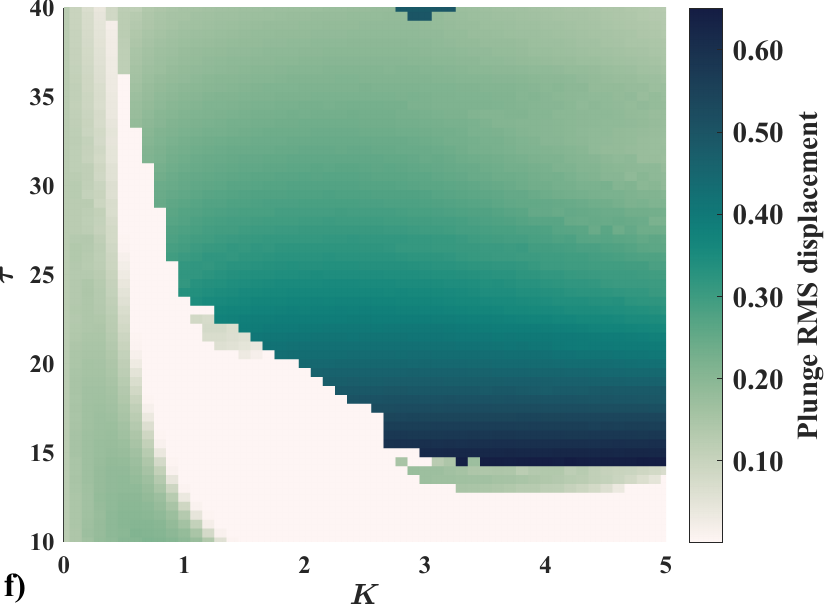}
	\caption{The RMS displacements of the coupled airfoil systems (\ref{eq:coupled-airfoil1-state-equation})-(\ref{eq:coupled-airfoil2-state-equation}) for the Case III ($\varrho=1.0$ and $\theta=0.5$) under different coupling strengths $K$ and time delays $\tau$. (a, b) $T=10$; (c, d) $T=40$; (e, f) $T=70$.}
	\label{fig:RMS-displacements-CaseIII-Ktau-T}
\end{figure}

\begin{figure}[!t]
	\centering
	\includegraphics[width=0.48\columnwidth]{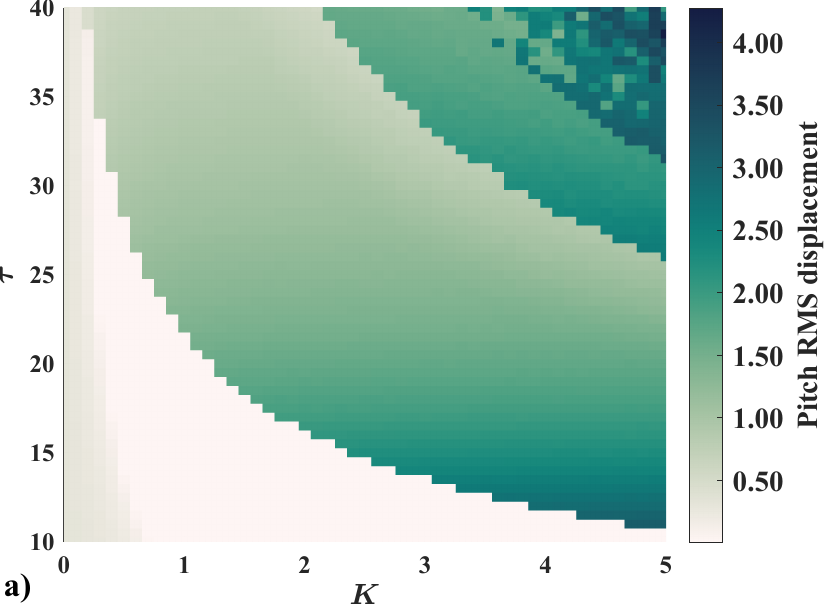} \quad
	\includegraphics[width=0.48\columnwidth]{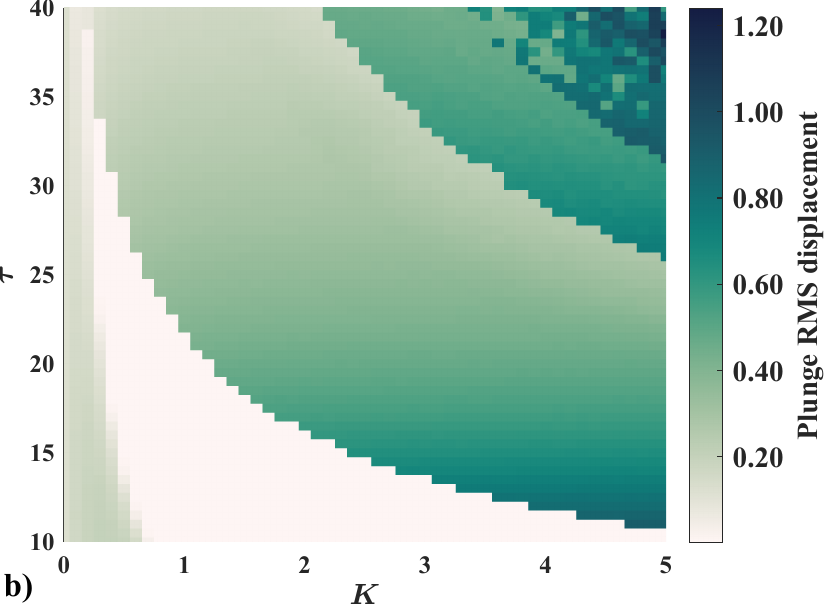} \\
	\includegraphics[width=0.48\columnwidth]{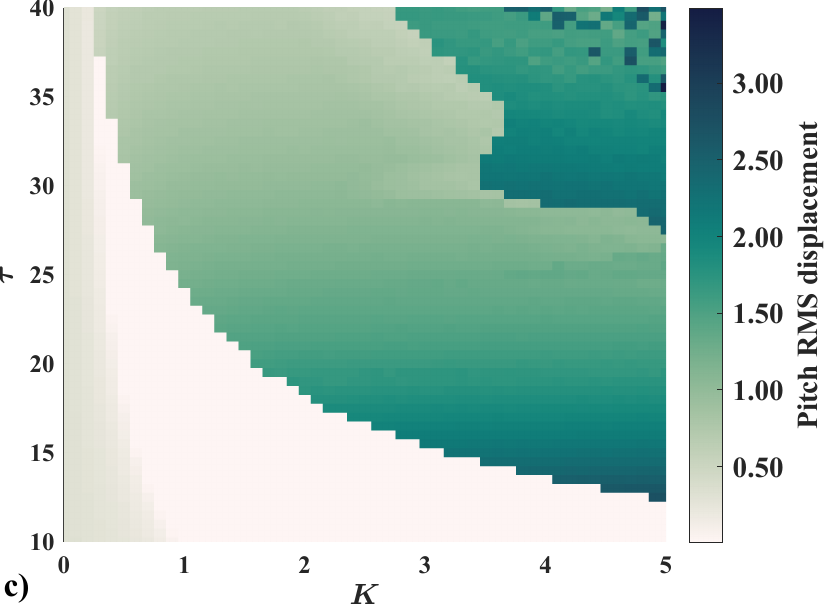} \quad
	\includegraphics[width=0.48\columnwidth]{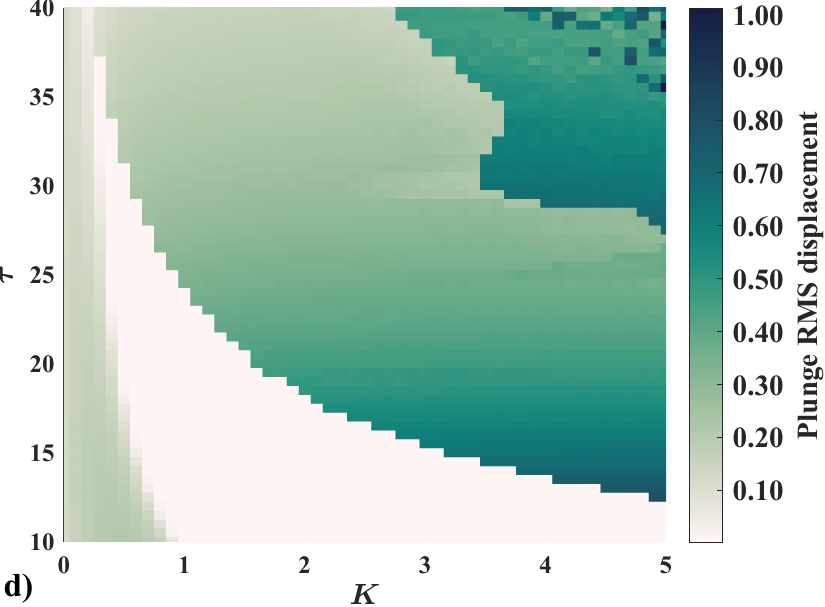} \\
	\includegraphics[width=0.48\columnwidth]{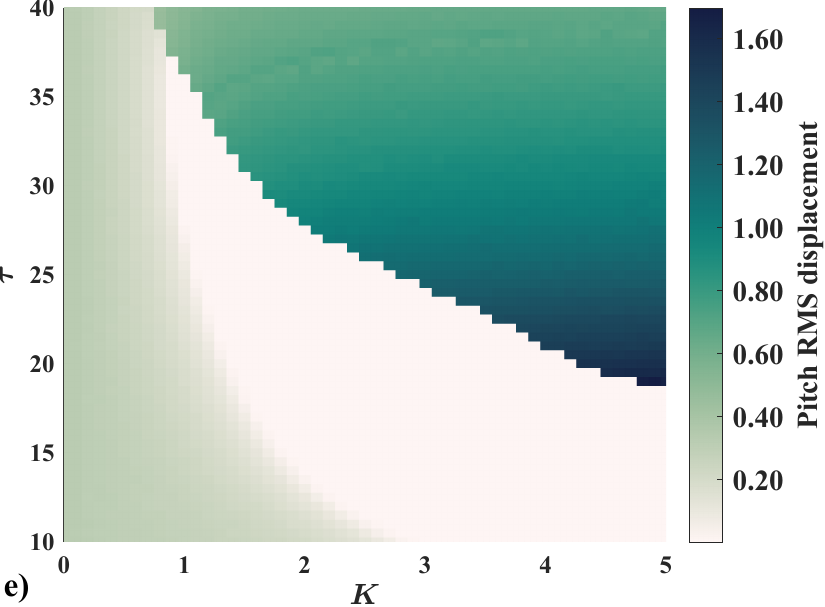} \quad
	\includegraphics[width=0.48\columnwidth]{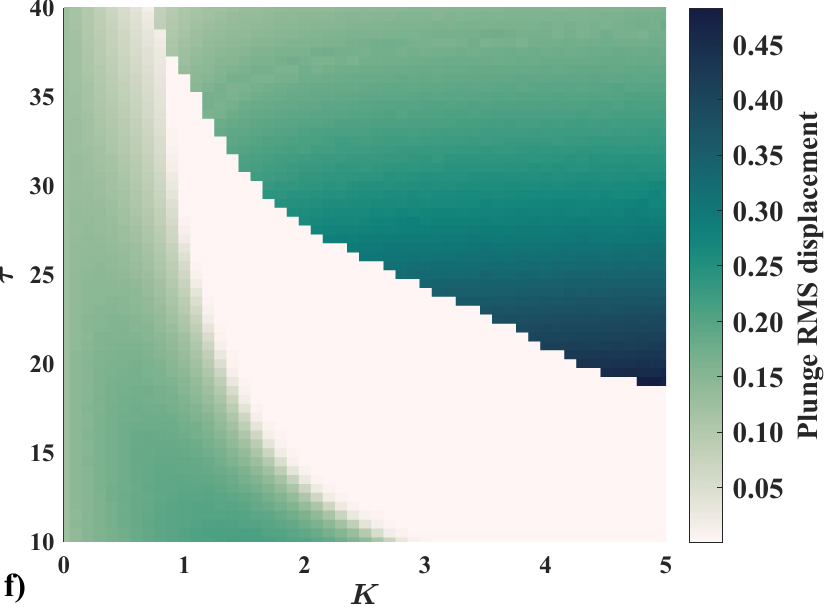}
	\caption{The RMS displacements of the coupled airfoil systems (\ref{eq:coupled-airfoil1-state-equation})-(\ref{eq:coupled-airfoil2-state-equation}) for the Case III ($\varrho=1.0$ and $T=10$) under different coupling strengths $K$ and time delays $\tau$. (a, b) $\theta=1$; (c, d) $\theta=0.75$; (e, f) $\theta=0.25$.}
	\label{fig:RMS-displacements-CaseIII-Ktau-theta}
\end{figure}

\begin{figure}[!t]
	\centering
	\includegraphics[width=0.48\columnwidth]{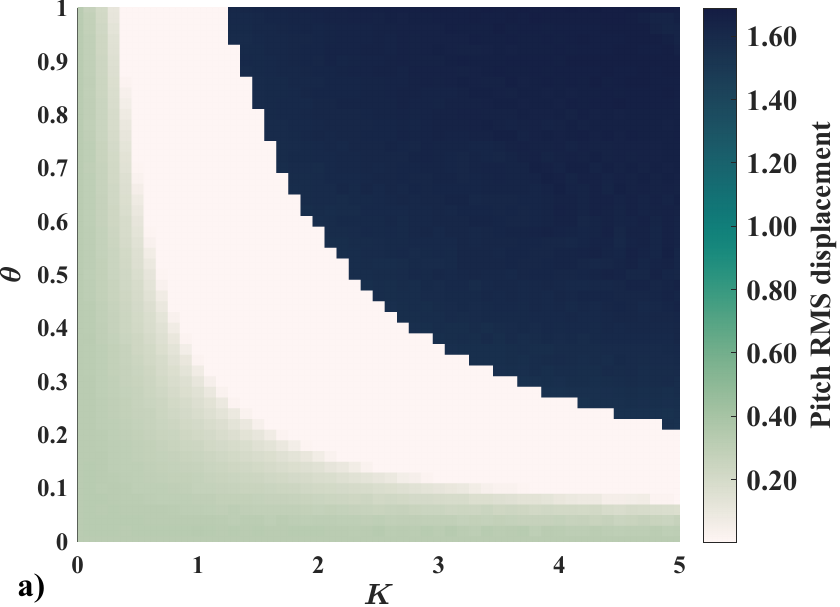} \quad
	\includegraphics[width=0.48\columnwidth]{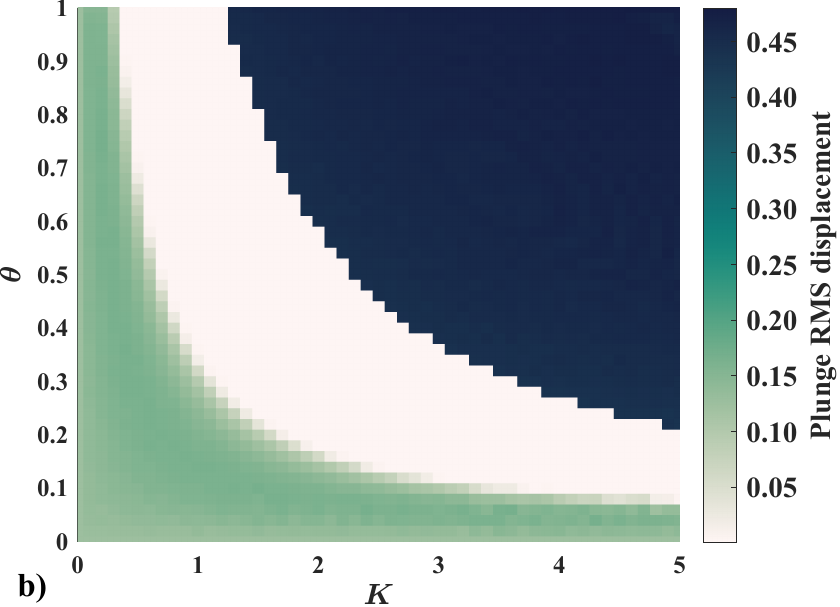} \\
	\includegraphics[width=0.48\columnwidth]{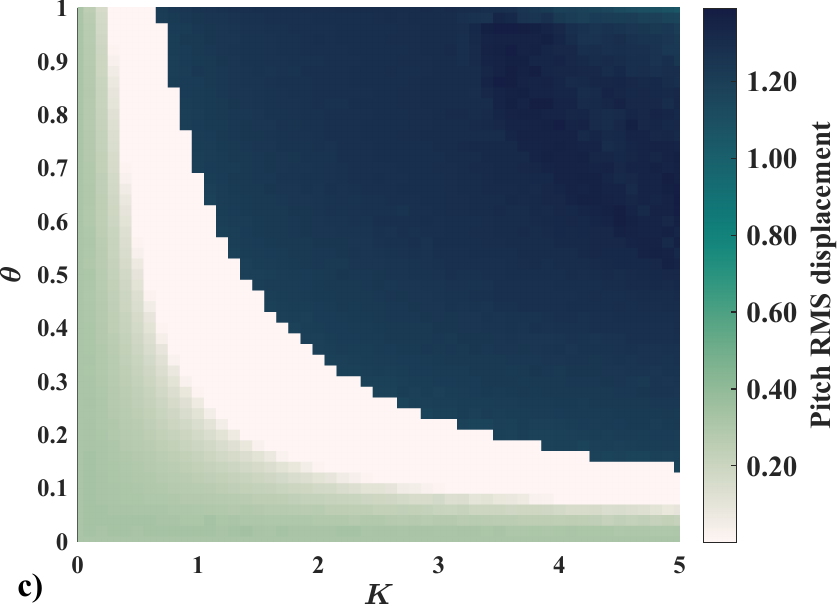} \quad
	\includegraphics[width=0.48\columnwidth]{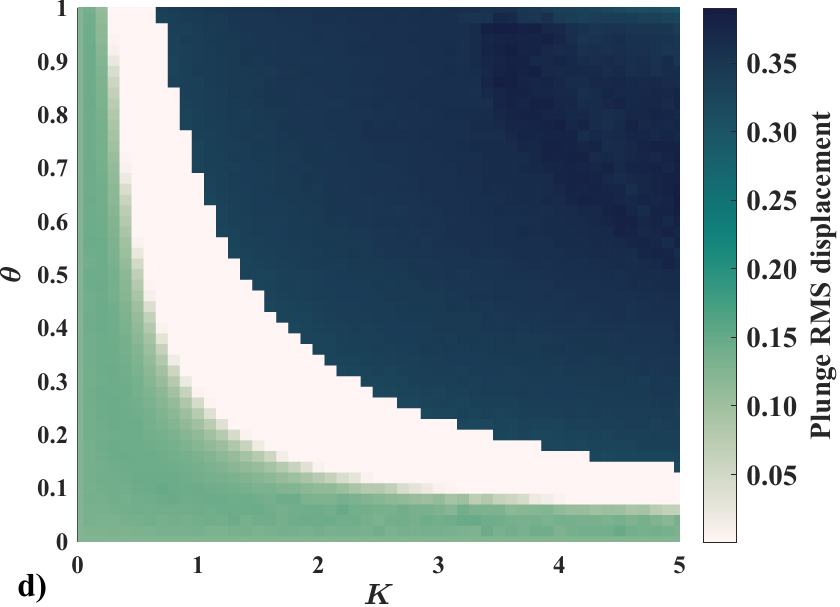} \\
	\includegraphics[width=0.48\columnwidth]{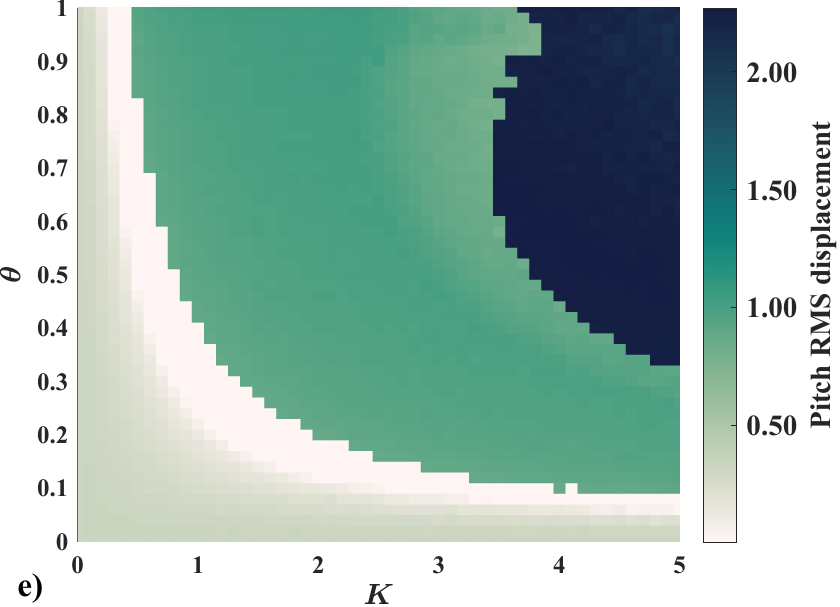} \quad
	\includegraphics[width=0.48\columnwidth]{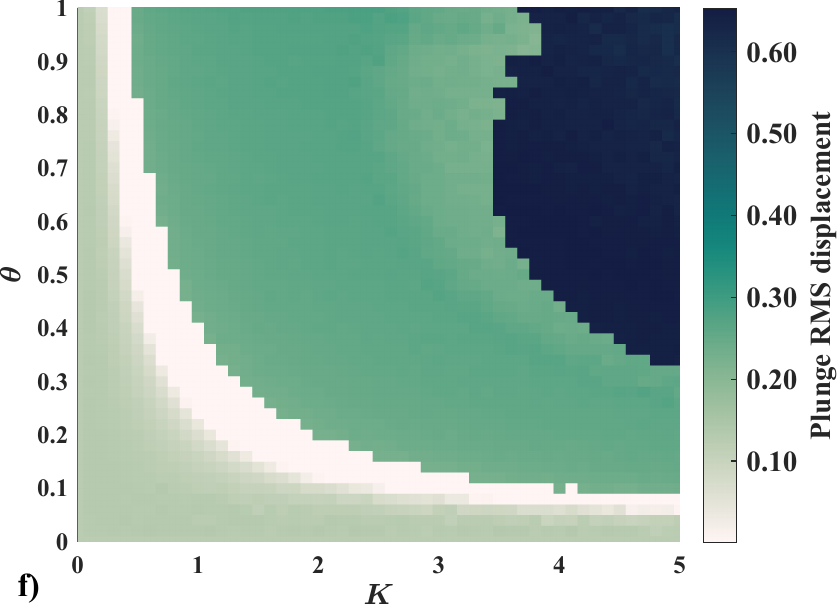}
	\caption{The RMS displacements of the coupled airfoil systems (\ref{eq:coupled-airfoil1-state-equation})-(\ref{eq:coupled-airfoil2-state-equation}) for the Case III ($\varrho=1.0$ and $T=10$) under different coupling strengths $K$ and parameters $\theta$. (a, b) $\tau=20$; (c, d) $\tau=25$; (e, f) $\tau=30$.}
	\label{fig:RMS-displacements-CaseIII-Ktheta-tau}
\end{figure}

\begin{figure}[!t]
	\centering
	\includegraphics[width=0.75\columnwidth]{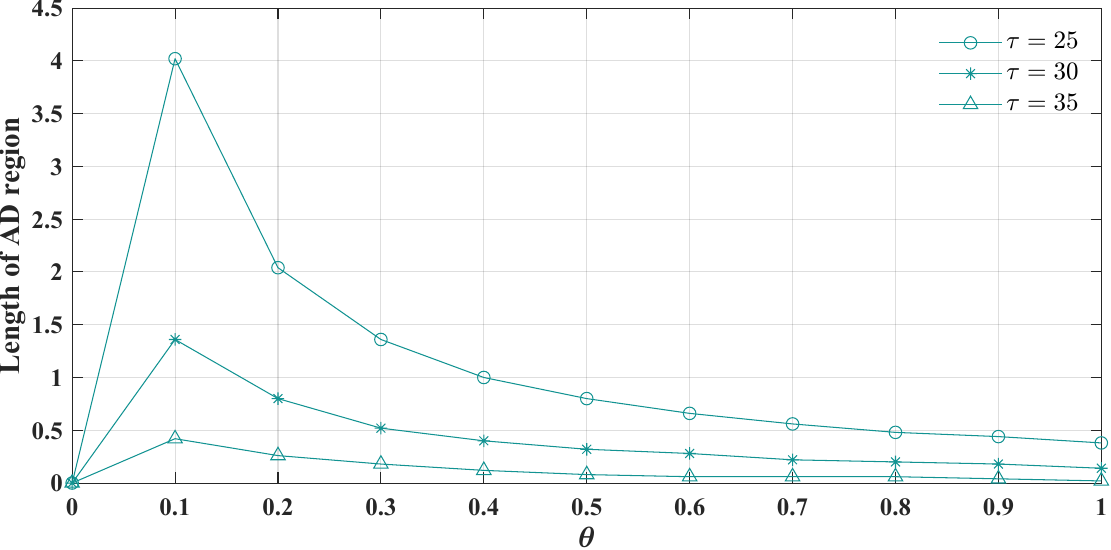}
	\caption{The length of AD region against the on-off ratio $\theta$ for the Case III with a fixed on-off period $T=10$.}
	\label{fig:length-of-ADregion-CaseIII-theta}
\end{figure}

Subsequently, we consider the effects of the intermittent purely time-delayed coupling, i.e., the parameter $\varrho=1$ and $\chi_{T,\theta} \left( t \right)$ varies periodically between 1 and 0. First, we discuss the effects of the on-off period $T$ at a fixed on-off ratio $\theta=0.5$ and plot the RMS displacements of the coupled airfoil systems (\ref{eq:coupled-airfoil1-state-equation})-(\ref{eq:coupled-airfoil2-state-equation}) for the Case III under different combinations of coupling strengths $K$ and time delay $\tau$, as shown in Fig.~\ref{fig:RMS-displacements-CaseIII-Ktau-T}. The results show that the choice of the parameter $T$ has significant effects on the emergence of AD. The AD region tends to decrease when $T$ increases. In addition, we find that the selection of the parameter $T$ is not directly related to the period $T_{\text{LCOs}}$ of LCOs. When the on-off period $T$ is close to $T_{\text{LCOs}}$ (see Figs.~\ref{fig:RMS-displacements-CaseIII-Ktau-T}e and \ref{fig:RMS-displacements-CaseIII-Ktau-T}f), the AD region becomes smaller. In the next analysis, we will fix the on-off period $T=10$, although it is not the optimal choice.

Then, we discuss the effects of another important parameter, i.e., the on-off ratio $\theta$ in the intermittent coupling for the Case III. We fix the on-off period $T=10$ and present the RMS displacements of the coupled airfoil systems (\ref{eq:coupled-airfoil1-state-equation})-(\ref{eq:coupled-airfoil2-state-equation}) for the on-off ratios $\theta=1$, $\theta=0.75$, and $\theta=0.25$, respectively. The results are presented in Fig.~\ref{fig:RMS-displacements-CaseIII-Ktau-theta}. Together with Figs.~\ref{fig:RMS-displacements-CaseIII-Ktau-T}a and \ref{fig:RMS-displacements-CaseIII-Ktau-T}b, we can confirm that the range of coupling strength $K$ where the coupled airfoil systems (\ref{eq:coupled-airfoil1-state-equation})-(\ref{eq:coupled-airfoil2-state-equation}) experience the AD behaviors becomes large as the on-off ratio $\theta$ decreases. In particular, compared to the continuous coupling, i.e., $\theta=1$, the presence of intermittent interactions can effectively expand the AD regions. To further visualize the effects of the on-off ratio $\theta$, we present the results in the parameter space ($K, \theta$) for different values of $\tau$, as depicted in Fig.~\ref{fig:RMS-displacements-CaseIII-Ktheta-tau}. The parameter space we consider is $(K, \theta)=[0,5] \times [0,1]$. The results show that the parameter range of coupling strengths $K$ where AD occurs increases as the on-off ratio $\theta$ decreases and the enhancement performance diminishes as the time delay $\tau$ increases. But the minimum value of coupling strength $K$ that can trigger the emergence of AD increases as $\theta$ decreases. The effects of the parameter $\theta$ in the intermittent coupling are analogous to those of the parameter $\varrho$ controlling the mixed coupling.

In order to quantify the size of AD region, we numerically computed the length of AD region corresponding to the length of the coupling strength parameter $K$ for which the AD behavior appears. We consider the numerical step size $\Delta K=0.02$ of the coupling strength $K$. The smaller $\Delta K$ is, the more accurate the calculation of the length of AD region is. The results on the variation of the length of AD region with the on-off ratio $\theta$ for three typical time delays, including $\tau=25$, $\tau=30$, and $\tau=35$, are given in Fig.~\ref{fig:length-of-ADregion-CaseIII-theta}. When $\theta=0$, the term $\chi_{T,\theta} \left( t \right)\equiv 0$, as shown in Fig.~\ref{fig:intermittent-coupling-signals}d, the coupled airfoil systems (\ref{eq:coupled-airfoil1-state-equation})-(\ref{eq:coupled-airfoil2-state-equation}) degenerates to the uncoupled case. In this case, the AD behavior does not appear in the whole parameter range of coupling strength $K$, so the length of AD region is always zero. When $\theta\in (0,1)$, the length of AD region decreases with the increasing of $\theta$. Meanwhile, the length of AD region in the range $\theta\in (0,1)$ is always larger than that in the case of $\theta=1$, i.e., the case of continuous coupling ($\chi_{T,\theta} \left( t \right)\equiv 1$), as shown in Fig.~\ref{fig:intermittent-coupling-signals}a), indicating that the introduction of intermittent interactions can expand the parameter range in which AD occurs and enhance the flutter suppression performance. Besides, the length of AD region decreases and the enhancement performance diminishes as the time delay $\tau$ increases. For example, for $\theta=0.1$, when $\tau=25$, the intermittent coupling causes the the length of AD region to increase by 3.64 relative to the continuous coupling case ($\theta=1$); however, when $\tau=35$, the intermittent coupling increases the length by only 0.4, although the relative increase in percentage may be large.

\subsubsection{Results for the Case IV}

\begin{figure}[!t]
	\centering
	\includegraphics[width=0.48\columnwidth]{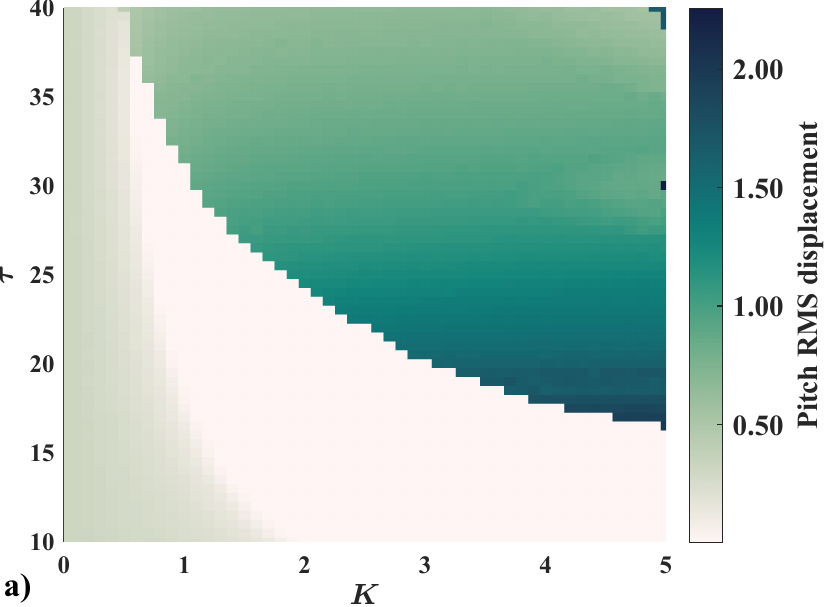} \quad
	\includegraphics[width=0.48\columnwidth]{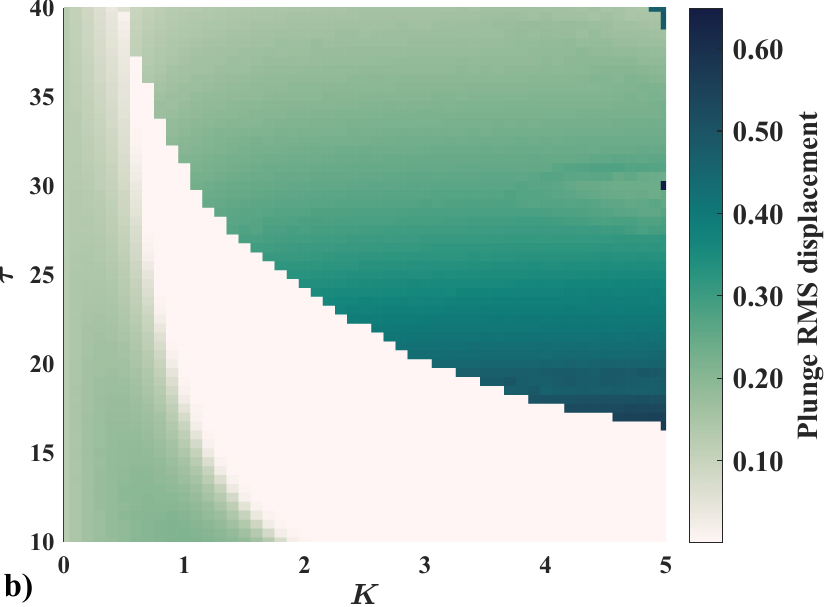} \\
	\includegraphics[width=0.48\columnwidth]{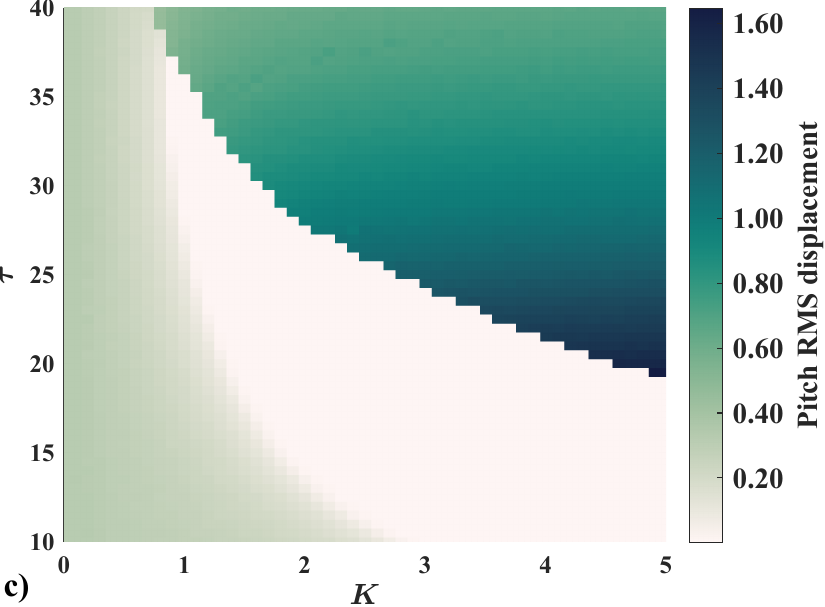} \quad
	\includegraphics[width=0.48\columnwidth]{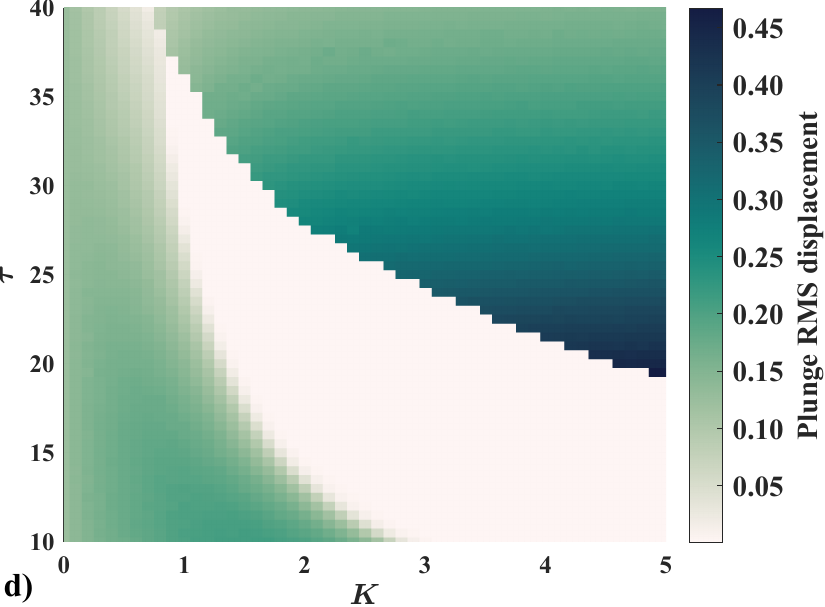} \\
	\includegraphics[width=0.48\columnwidth]{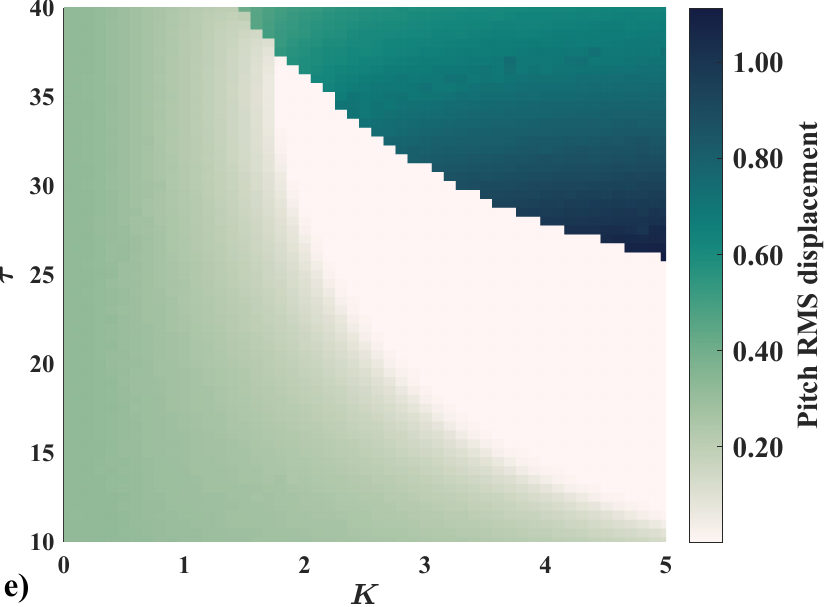} \quad
	\includegraphics[width=0.48\columnwidth]{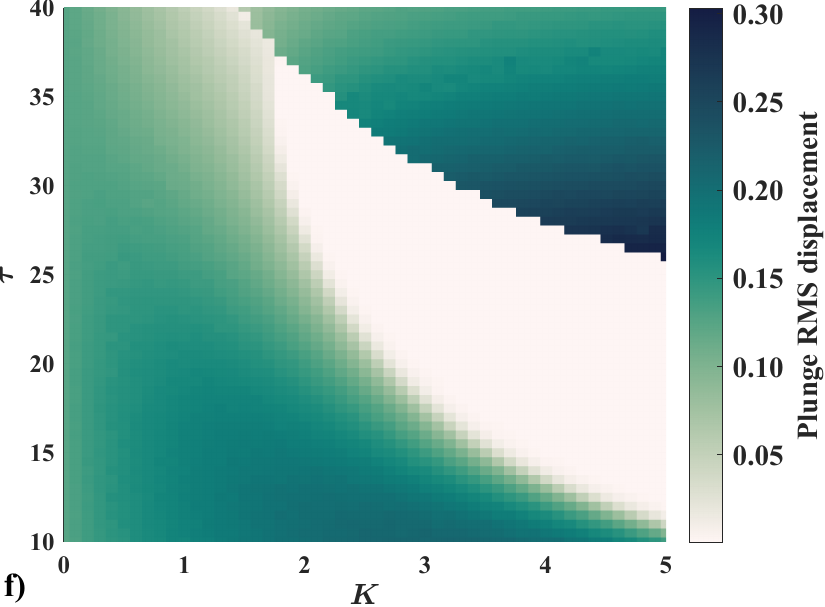}
	\caption{The RMS displacements of the coupled airfoil systems (\ref{eq:coupled-airfoil1-state-equation})-(\ref{eq:coupled-airfoil2-state-equation}) for the Case IV ($T=10$ and $\theta=0.5$) under different coupling strengths $K$ and time delays $\tau$. (a, b) $\varrho=0.75$; (c, d) $\varrho=0.5$; (e, f) $\varrho=0.25$.}
	\label{fig:RMS-displacements-CaseIV-Ktau-varrho}
\end{figure}

\begin{figure}[!t]
	\centering
	\includegraphics[width=0.48\columnwidth]{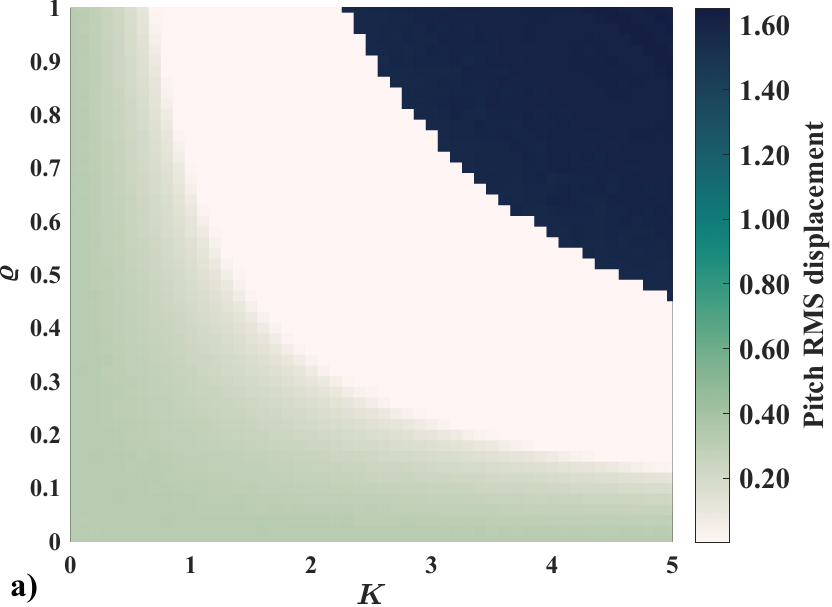} \quad
	\includegraphics[width=0.48\columnwidth]{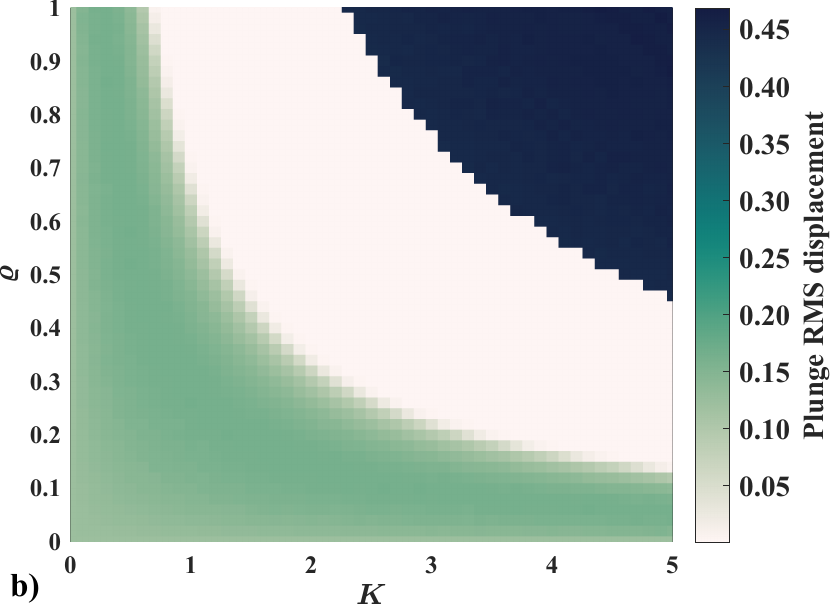} \\
	\includegraphics[width=0.48\columnwidth]{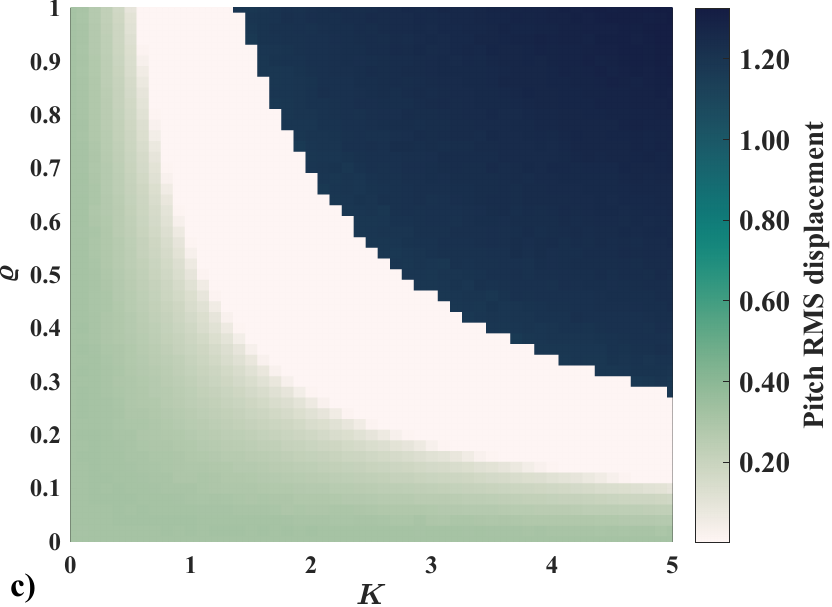} \quad
	\includegraphics[width=0.48\columnwidth]{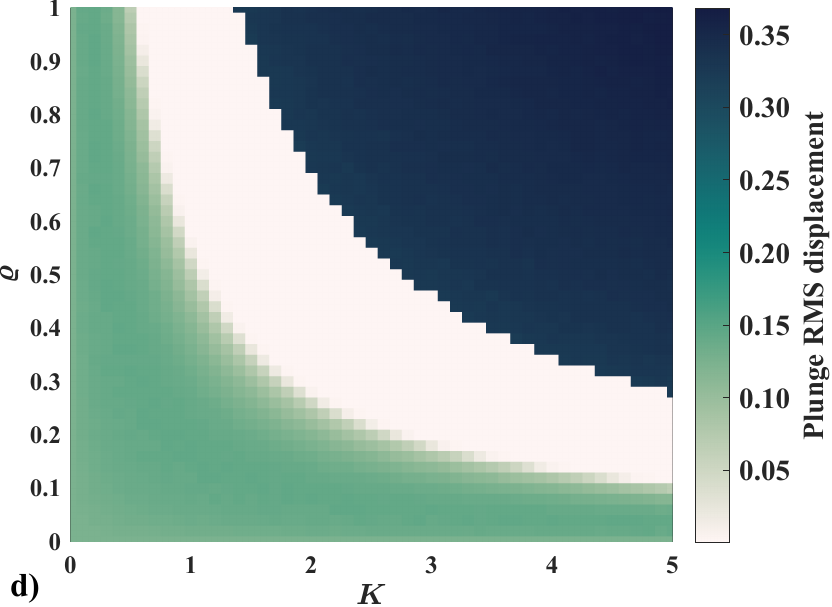} \\
	\includegraphics[width=0.48\columnwidth]{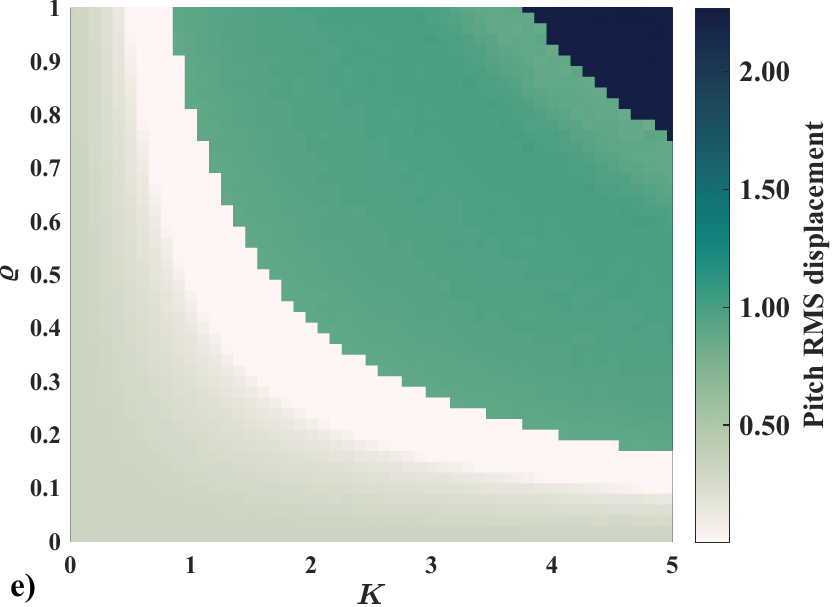} \quad
	\includegraphics[width=0.48\columnwidth]{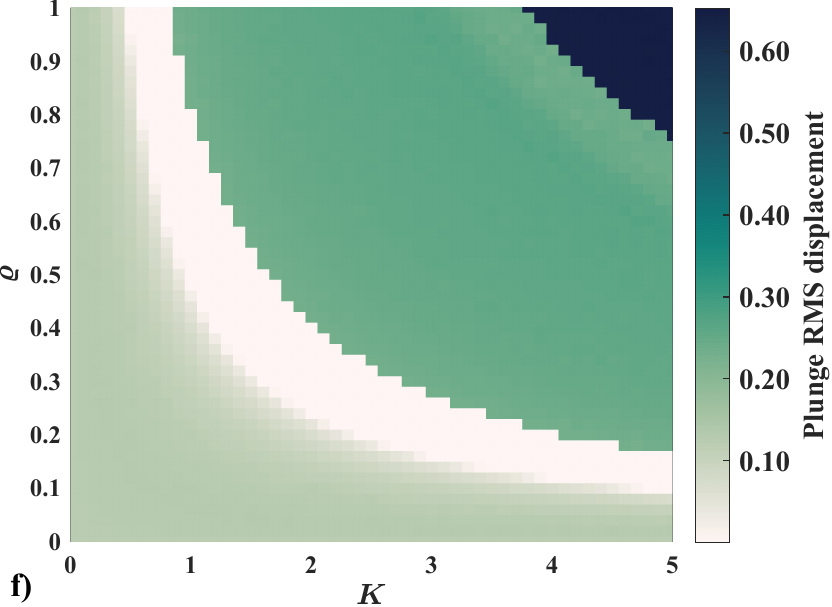}
	\caption{The RMS displacements of the coupled airfoil systems (\ref{eq:coupled-airfoil1-state-equation})-(\ref{eq:coupled-airfoil2-state-equation}) for the Case IV ($T=10$ and $\theta=0.5$) under different coupling strengths $K$ and parameters $\varrho$. (a, b) $\tau=20$; (c, d) $\tau=25$; (e, f) $\tau=30$.}
	\label{fig:RMS-displacements-CaseIV-Kvarrho-tau}
\end{figure}

\begin{figure}[!t]
	\centering
	\includegraphics[width=0.48\columnwidth]{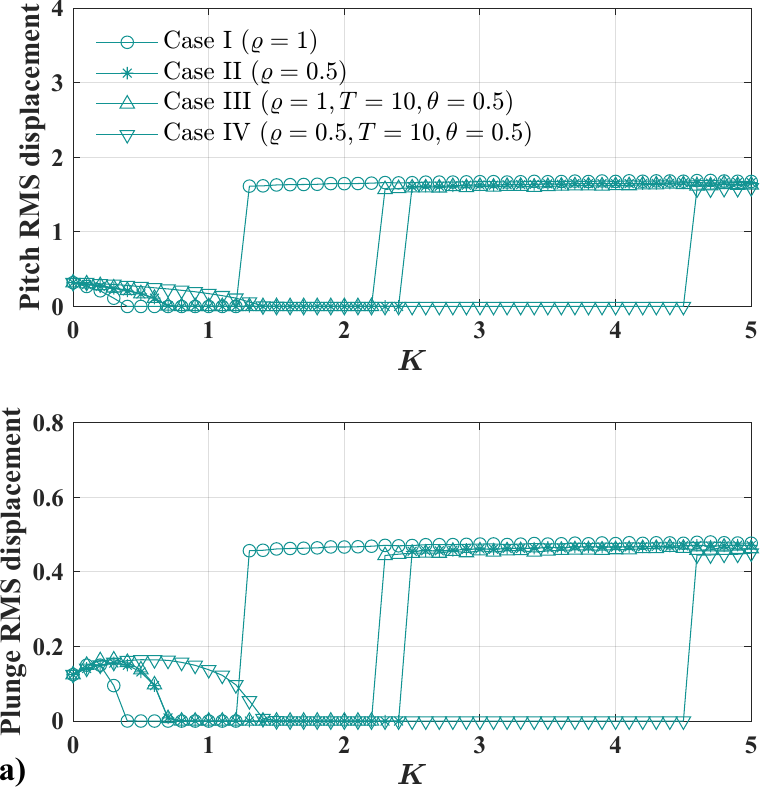} \quad
	\includegraphics[width=0.48\columnwidth]{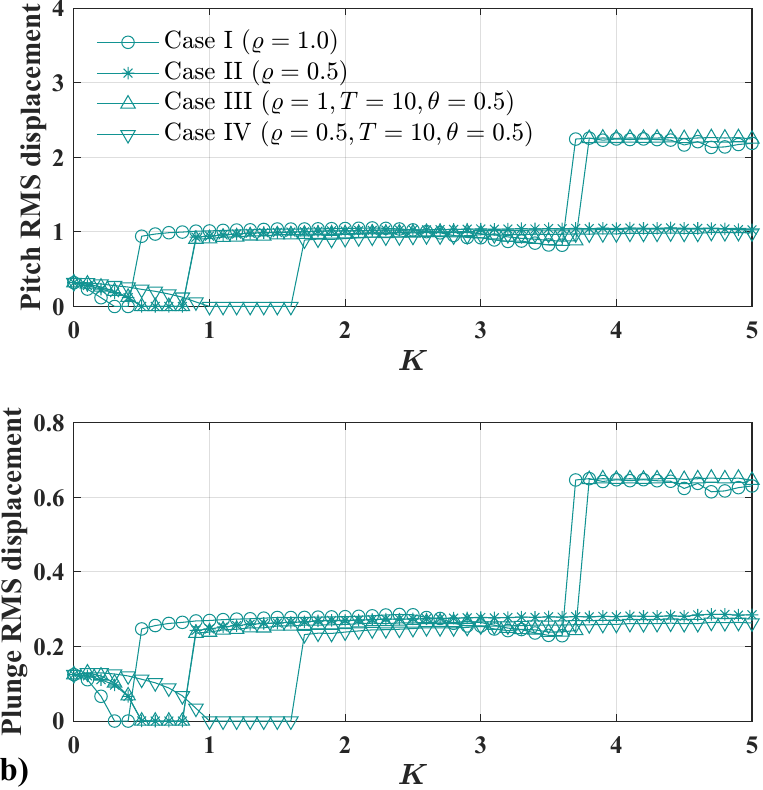}
	\caption{Comparison results of the RMS displacements of the coupled airfoil systems (\ref{eq:coupled-airfoil1-state-equation})-(\ref{eq:coupled-airfoil2-state-equation}) for different cases under different time delay $\tau$. (a) $\tau=20$; (b) $\tau=30$.}
	\label{fig:RMS-displacements-CaseIV-K-tau}
\end{figure}

\begin{figure}[!t]
	\centering
	\includegraphics[width=0.75\columnwidth]{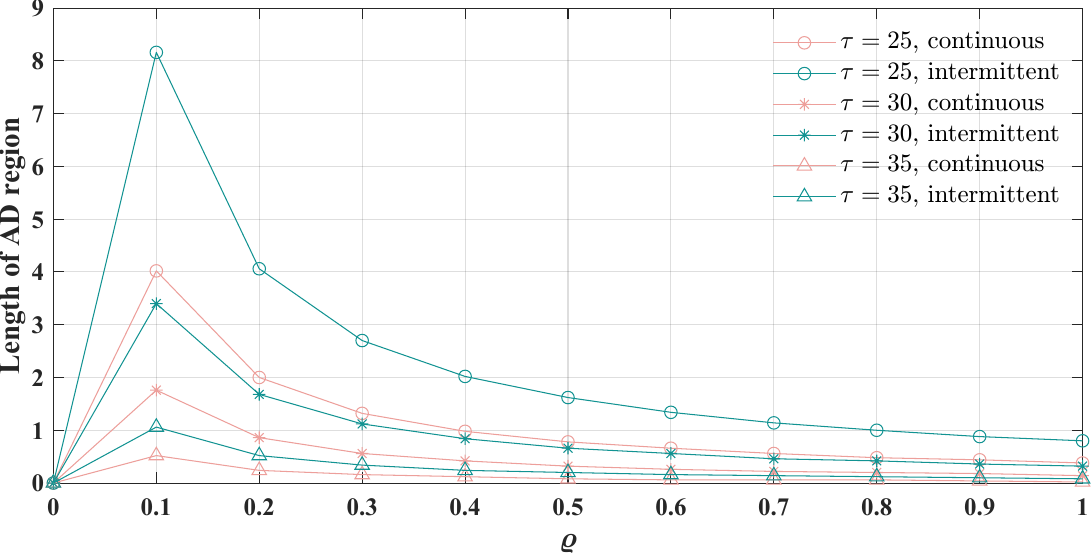}
	\caption{The length of AD region against the parameter $\varrho$.}
	\label{fig:length-of-ADregion-CaseIV-varrho}
\end{figure}

Finally, we examine the case where the intermittent mixed coupling, i.e., the combined effect of both intermittent instantaneous coupling and intermittent time-delayed coupling, are present in the coupled airfoil systems (\ref{eq:coupled-airfoil1-state-equation})-(\ref{eq:coupled-airfoil2-state-equation}). In this case, the parameter $\varrho\in (0,1)$ and $\chi_{T,\theta} \left( t \right)$ varies periodically between 1 and 0. We fix the on-off period $T=10$ and the on-off ratio $\theta=0.5$, and explore the influences of the parameter $\varrho$ on the systems' dynamics. The results are indicated in Figs.~\ref{fig:RMS-displacements-CaseIV-Ktau-varrho} and \ref{fig:RMS-displacements-CaseIV-Kvarrho-tau}, which display the RMS displacements in the parameter space $(K, \tau)$ for different $\varrho$ and in the parameter space ($K, \varrho$) for different $\tau$, respectively. Compared with the results of the continuous mixed coupling cases (Case II), i.e., Figs.~\ref{fig:RMS-displacements-CaseII-Ktau-varrho} and \ref{fig:RMS-displacements-CaseII-Kvarrho-tau}, we observe that the introduction of intermittent interactions between two airfoils can further enhance the AD regime, i.e., enlarging the parameter domains where AD occurs, but also increases the minimum coupling strength that triggers the AD behaviors.

To characterize more clearly the advantages of the coupling strategy proposed in this study, we present a comparison of the results for the four coupling scenarios, i.e., Case I--Case IV. The results for two given time delays $\tau=20$ and $\tau=30$ are shown in Fig.~\ref{fig:RMS-displacements-CaseIV-K-tau}. We also show the variation of the length of the AD region with the parameter $\varrho$ for the continuous coupling (light red curves) and the intermittent coupling (green curves) cases, respectively. The results for a fixed on-off period $T=10$ are shown in Fig.~\ref{fig:length-of-ADregion-CaseIV-varrho}. The results reveal that in the continuous coupling case, the introduction of the continuous mixed coupling (Case II) can enhance the AD regime compared to the continuous purely time-delayed coupling (Case I). At the same time, the introduction of the intermittent interactions (Case III and Case IV) can realize the further enhancement of the AD regimes in both Case I and Case II. Besides, from Fig.~\ref{fig:length-of-ADregion-CaseIV-varrho}, it can be noticed that the effect of parameter $\rho$ is similar to that of parameter $\theta$ (see Fig.~\ref{fig:length-of-ADregion-CaseIII-theta}). When $\varrho=0$, the coupling degenerates to purely instantaneous coupling between two identical airfoils and does not yield AD, which corresponds to the zero length of the AD region. For the mixed coupling case ($\varrho\in (0,1)$), the length decreases gradually with increasing $\varrho$, and all of them are larger than the purely time-delayed coupling case ($\varrho=1$). More importantly, when $\varrho$ is fixed, the intermittent coupling can further extend the length of AD region compared to the continuous coupling case. The enhancement effect of the intermittent coupling in the small time-delayed case will be more obvious than the large time-delayed case.

\section{Conclusions} \label{sec:sec-4}

In the present study, the flutter suppression enhancement strategy for two coupled identical airfoils with cubic structural nonlinearity was explored based on the AD mechanism. Intermittent mixed interactions between two airfoils were considered to enlarge the AD regime to realize the flutter suppression enhancement. The influence mechanisms of different coupling forms on the dynamical behaviors of the coupled airfoil systems were discussed in detail through direct numerical simulations. We showed that the two coupled identical airfoils will exhibit the AD behaviors within a certain range of coupling strength and time delay. In particular, the mixed coupling strategy favors the AD phenomenon over a larger parameter set of the coupling strength than the limited case of purely time-delayed coupling. Moreover, in comparison with the continuous coupling setting, the intermittent coupling leads to further expansion of the expected AD region and can achieve more efficiently the flutter suppression enhancement. All the findings provide new insights into the design of aircraft wing structures to achieve the flutter suppression enhancement and ensure the flight safety and structural integrity of an aircraft.

We have showed the impacts of different coupling strategies from theoretical modeling and numerical perspectives but have not taken into account their physical significance in practice. Future works could explore the practical realization of intermittent mixed coupling, and the flutter suppression of the coupled airfoils without time-delayed interactions via considering parameter mismatch \cite{raj2021effect} or other strategies like dynamic coupling \cite{konishi2003amplitude} and nonlinear coupling \cite{prasad2010amplitude}. Moreover, the above strategies could be further enriched by considering different kinds of interactions between the airfoils, such as higher-order (many-body) \cite{battiston2021physics,millan2025topology} or non-normal ones \cite{trefethen2005,muolo2021synchronization}, both known to have great effects on the dynamics and, in particular, on synchronization dynamics. Lastly, let us point out that our results go beyond the framework of an aircraft and can pave the way in further studies of AD both from a purely theoretical perspective and in applications regarding the dynamics of coupled oscillators such as wind turbine blades and power grids.

\section*{Appendix A: State equations of the two coupled airfoils}

Regarding the governing equations (\ref{eq:coupled-airfoil1-plunge})-(\ref{eq:coupled-airfoil2-pitch}) of the two coupled airfoils, we introduce the following variables
\begin{equation*}
	w_{i1}=\int_0^{t}{e^{-\varepsilon _1\left( t -\sigma \right)}\alpha_{i} \left( \sigma \right) \text{d}\sigma},\quad w_{i2}=\int_0^{t}{e^{-\varepsilon _2\left( t -\sigma \right)}\alpha_{i} \left( \sigma \right) \text{d}\sigma},
\end{equation*}
\begin{equation*}
	w_{i3}=\int_0^{t}{e^{-\varepsilon _1\left( t -\sigma \right)}\xi_{i} \left( \sigma \right) \text{d}\sigma},\quad w_{i4}=\int_0^{t}{e^{-\varepsilon _2\left( t -\sigma \right)}\xi_{i} \left( \sigma \right) \text{d}\sigma},
\end{equation*}
where $i=1, 2$ indicate the first and second airfoil systems, respectively. Denoting the state vector of the $i$-th airfoil system as $\boldsymbol{x}_{i}=\left( x_{i1},x_{i2},...,x_{i8} \right) ^{\text{T}} =\left( \alpha_{i}, \dot{\alpha}_{i}, \xi_{i}, \dot{\xi}_{i}, w_{i1}, w_{i2}, w_{i3}, w_{i4} \right) ^{\text{T}} \in \mathbb{R}^8$, Eqs.~(\ref{eq:coupled-airfoil1-plunge})-(\ref{eq:coupled-airfoil2-pitch}) then will be expressed as a series of first-order ordinary differential equations
\begin{equation*}
	\left\{ \begin{array}{l}
		\dot{x}_{11}=x_{12},\\
		\dot{x}_{12}=a_{21}x_{11}+a_{22}x_{12}+a_{23}x_{13}+a_{24}x_{14}+a_{25}x_{15}+a_{26}x_{16}+a_{27}x_{17}+a_{28}x_{18}+\left[ g_{23}G\left( x_{13} \right) - g_{21}M\left( x_{11} \right) \right]\\
		\qquad\quad -\frac{c_0}{D} \left[ \left( 1 - \varrho \right) \cdot \chi_{T,\theta} \left( t \right) \cdot \frac{K}{U^{*2}} \left( x_{11}-x_{21} \right) + \varrho \cdot \chi_{T,\theta} \left( t \right) \cdot \frac{K}{U^{*2}} \left( x_{11}-x_{21}\left( t -\tau \right) \right) \right],\\
		\dot{x}_{13}=x_{14},\\
		\dot{x}_{14}=a_{41}x_{11}+a_{42}x_{12}+a_{43}x_{13}+a_{44}x_{14}+a_{45}x_{15}+a_{46}x_{16}+a_{47}x_{17}+a_{48}x_{18}+\left[ g_{41}M\left( x_{11} \right) - g_{43}G\left( x_{13} \right) \right]\\
		\qquad\quad +\frac{c_1}{D} \left[ \left( 1 - \varrho \right) \cdot \chi_{T,\theta} \left( t \right) \cdot \frac{K}{U^{*2}} \left( x_{11}-x_{21} \right) + \varrho \cdot \chi_{T,\theta} \left( t \right) \cdot \frac{K}{U^{*2}} \left( x_{11}-x_{21}\left( t -\tau \right) \right) \right],\\
		\dot{x}_{15}=x_{11}-\varepsilon _1x_{15},\\
		\dot{x}_{16}=x_{11}-\varepsilon _2x_{16},\\
		\dot{x}_{17}=x_{13}-\varepsilon _1x_{17},\\
		\dot{x}_{18}=x_{13}-\varepsilon _2x_{18},
	\end{array} \right.
\end{equation*}
and
\begin{equation*}
	\left\{ \begin{array}{l}
		\dot{x}_{21}=x_{22},\\
		\dot{x}_{22}=a_{21}x_{21}+a_{22}x_{22}+a_{23}x_{23}+a_{24}x_{24}+a_{25}x_{25}+a_{26}x_{26}+a_{27}x_{27}+a_{28}x_{28}+\left[ g_{23}G\left( x_{23} \right) - g_{21}M\left( x_{21} \right) \right]\\
		\qquad\quad -\frac{c_0}{D} \left[ \left( 1 - \varrho \right) \cdot \chi_{T,\theta} \left( t \right) \cdot \frac{K}{U^{*2}} \left( x_{21}-x_{11} \right) + \varrho \cdot \chi_{T,\theta} \left( t \right) \cdot \frac{K}{U^{*2}} \left( x_{21}-x_{11}\left( t -\tau \right) \right) \right],\\
		\dot{x}_{23}=x_{24},\\
		\dot{x}_{24}=a_{41}x_{21}+a_{42}x_{22}+a_{43}x_{23}+a_{44}x_{24}+a_{45}x_{25}+a_{46}x_{26}+a_{47}x_{27}+a_{48}x_{28}+\left[ g_{41}M\left( x_{21} \right) - g_{43}G\left( x_{23} \right) \right]\\
		\qquad\quad +\frac{c_1}{D} \left[ \left( 1 - \varrho \right) \cdot \chi_{T,\theta} \left( t \right) \cdot \frac{K}{U^{*2}} \left( x_{21}-x_{11} \right) + \varrho \cdot \chi_{T,\theta} \left( t \right) \cdot \frac{K}{U^{*2}} \left( x_{21}-x_{11}\left( t -\tau \right) \right) \right],\\
		\dot{x}_{25}=x_{21}-\varepsilon _1x_{25},\\
		\dot{x}_{26}=x_{21}-\varepsilon _2x_{26},\\
		\dot{x}_{27}=x_{23}-\varepsilon _1x_{27},\\
		\dot{x}_{28}=x_{23}-\varepsilon _2x_{28},
	\end{array} \right.
\end{equation*}
where the specific expressions of the coefficients $a_{ij}$ ($i=2,4; j=1,2,...,8$) and $g_{lk}$ ($l=2,4; k=1,3$) are detailed in Appendix B. Finally, we can obtain the state equations (\ref{eq:coupled-airfoil1-state-equation}) and (\ref{eq:coupled-airfoil2-state-equation}).

\section*{Appendix B: Coefficients in the state equations}
The coefficients in the state equations given in Appendix A are given as follows \cite{lee2006bifurcation}
\begin{align*}
	a_{21}&=\frac{c_5d_0-c_0d_5}{D}, \quad a_{22}=\frac{c_3d_0-c_0d_3}{D}, \quad a_{23}=\frac{c_4d_0-c_0d_4}{D}, \quad a_{24}=\frac{c_2d_0-c_0d_2}{D},\\ a_{25}&=\frac{c_6d_0-c_0d_6}{D}, \quad a_{26}=\frac{c_7d_0-c_0d_7}{D}, \quad
	a_{27}=\frac{c_8d_0-c_0d_8}{D}, \quad a_{28}=\frac{c_9d_0-c_0d_9}{D},\\
	a_{41}&=\frac{c_1d_5-c_5d_1}{D}, \quad a_{42}=\frac{c_1d_3-c_3d_1}{D}, \quad a_{43}=\frac{c_1d_4-c_4d_1}{D}, \quad
	a_{44}=\frac{c_1d_2-c_2d_1}{D},\\ a_{45}&=\frac{c_1d_6-c_6d_1}{D}, \quad a_{46}=\frac{c_1d_7-c_7d_1}{D}, \quad
	a_{47}=\frac{c_1d_8-c_8d_1}{D}, \quad a_{48}=\frac{c_1d_9-c_9d_1}{D},\\
	g_{21}&=\frac{c_0d_{10}}{D}, \quad
	g_{23}=\frac{c_{10}d_{0}}{D}, \quad g_{41}=\frac{c_1d_{10}}{D}, \quad
	g_{43}=\frac{c_{10}d_{1}}{D},
\end{align*}
where $D=c_0d_1-c_1d_0$ and
\begin{align*}
	c_0&=1+\frac{1}{\mu} ,\quad c_1=x_{\alpha}-\frac{a_h}{\mu},\quad c_2=\frac{2}{\mu}\left( 1-\psi _1-\psi _2 \right) +2\zeta _{\xi}\frac{\overline{\omega}}{U^*},\quad c_3=\frac{1}{\mu}\left[ 1+\left( 1-2a_h \right) \left( 1-\psi _1-\psi _2 \right) \right],\\
	c_4&=\frac{2}{\mu}\left( \varepsilon _1\psi _1+\varepsilon _2\psi _2 \right),\quad c_5=\frac{2}{\mu}\left[ 1-\psi _1-\psi _2+\left( \frac{1}{2}-\alpha _h \right) \left( \varepsilon _1\psi _1+\varepsilon _2\psi _2 \right) \right],\quad c_6=\frac{2}{\mu}\varepsilon _1\psi _1 \left[ 1-\varepsilon _1\left( \frac{1}{2}-a_h \right) \right],\\
	c_7&=\frac{2}{\mu}\varepsilon _2\psi _2\left[ 1-\varepsilon _2\left( \frac{1}{2}-\alpha _h \right) \right],\quad c_8=-\frac{2}{\mu}\varepsilon _{1}^{2}\psi _1,\quad c_9=-\frac{2}{\mu}\varepsilon _{2}^{2}\psi _2,\quad c_{10}=\left( \frac{\overline{\omega}}{U^*} \right) ^2,\\
	d_0&=\frac{x_{\alpha}}{r_{\alpha}^{2}}-\frac{a_h}{\mu r_{\alpha}^{2}},\quad d_1=1+\frac{1}{8\mu r_{\alpha}^{2}} \left( 1+8a_{h}^{2} \right),\quad d_2=-\frac{1}{\mu r_{\alpha}^{2}} \left( 1+2a_h \right) \left( 1-\psi _1-\psi _2 \right),\\
	d_3&=\frac{1}{2\mu r_{\alpha}^{2}}\left( 1-2a_h \right)-\frac{1}{2\mu r_{\alpha}^{2}} \left( 1+2a_h \right) \left( 1-2a_h \right) \left( 1-\psi _1-\psi _2 \right)+\frac{2\zeta _{\alpha}}{U^*},\\
	d_4&=-\frac{1}{\mu r_{\alpha}^{2}} \left( 1+2a_h \right) \left( \varepsilon _1\psi _1+\varepsilon _2\psi _2 \right),\quad d_5=-\frac{1}{\mu r_{\alpha}^{2}} \left( 1+2a_h \right) \left( 1-\psi _1-\psi _2 \right) -\frac{1}{2\mu r_{\alpha}^{2}} \left( 1+2a_h \right) \left( 1-2a_h \right) \left( \psi _1\varepsilon _1-\psi _2\varepsilon _2 \right),\\
	d_6&=-\frac{1}{\mu r_{\alpha}^{2}} \left( 1+2a_h \right) \psi _1\varepsilon _1\left[ 1-\varepsilon _1\left( \frac{1}{2}-a_h \right) \right],\quad d_7=-\frac{1}{\mu r_{\alpha}^{2}} \left( 1+2a_h \right) \psi _2\varepsilon _2\left[ 1-\varepsilon _2\left( \frac{1}{2}-a_h \right) \right],\\
	d_8&=\frac{1}{\mu r_{\alpha}^{2}} \left( 1+2a_h \right) \psi _1\varepsilon _{1}^{2},\quad d_9=\frac{1}{\mu r_{\alpha}^{2}} \left( 1+2a_h \right) \psi _2\varepsilon _{2}^{2},\quad d_{10}=\left( \frac{1}{U^*} \right) ^2.
\end{align*}

\section*{Acknowledgments}
This research was financially supported by the National Natural Science Foundation of China (Grant No.~52225211). Q.L. gratefully acknowledges the support of the China Scholarship Council (Grant No.~202006290313). The work of R.M. is supported by a JSPS postdoctoral fellowship, grant 24KF0211. H.N. acknowledges JSPS KAKENHI 25H01468, 25K03081, and 22H00516 for financial support.


\end{document}